\documentclass[10pt,journal,compsoc]{IEEEtran}
\usepackage{url}
\usepackage{graphicx}
\usepackage{subfigure}
\usepackage{booktabs}
\usepackage{bbding}
\usepackage{comment} 
\usepackage{verbatim}
\usepackage{multirow}
\usepackage{tabularx}
\usepackage{makecell}
\usepackage{diagbox}
\graphicspath{{./figures/}}

\usepackage{soul}
\usepackage{float}
\usepackage{enumitem}  
\usepackage{bbding}

\usepackage{cite}

\ifCLASSINFOpdf
\else
\fi
\usepackage{amsmath, amssymb}

\usepackage{xcolor}
\usepackage{tablefootnote}
\usepackage{soul}
\usepackage[flushleft]{threeparttable}

\usepackage{txfonts}
\usepackage{ifsym}

\begin{document}
%
\title{Adversarial Attacks Against Deep Generative Models on Data: A Survey}
%
%
%
%

\author{Hui Sun,
        Zhu Tianqing,~\IEEEmembership{Member,~IEEE,}
        Zhiqiu Zhang,
        Dawei Jin,~\IEEEmembership{Member,~IEEE,}
        Ping Xiong,~\IEEEmembership{Member,~IEEE,}
        and Wanlei Zhou,~\IEEEmembership{Senior Member,~IEEE}
\IEEEcompsocitemizethanks{\IEEEcompsocthanksitem Hui Sun, Zhu Tianqing, Zhiqiu Zhang are with the China University of Geosciences, Wuhan, China; Dawei Jin and Ping Xiong are with the Zhongnan University of Economy and Law, China; Wanlei Zhou is with the City University of Macau.\protect\\
\IEEEcompsocthanksitem Zhu Tianqing is the corresponding author. E-mail: tianqing.zhu@ieee.org}
}

%
%

\markboth{Journal of \LaTeX\ Class Files,~Vol.~14, No.~8, August~2015}%
{Shell \MakeLowercase{\textit{et al.}}: Bare Advanced Demo of IEEEtran.cls for IEEE Computer Society Journals}
%



\IEEEtitleabstractindextext{%
\begin{abstract}
Deep generative models have gained much attention given their ability to generate data for applications as varied as healthcare to financial technology to surveillance, and many more - the most popular models being generative adversarial networks (GANs) and variational auto-encoders (VAEs). Yet, as with all machine learning models, ever is the concern over security breaches and privacy leaks and deep generative models are no exception. In fact, these models have advanced so rapidly in recent years that work on their security is still in its infancy. In an attempt to audit the current and future threats against these models, and to provide a roadmap for defense preparations in the short term, we prepared this comprehensive and specialized survey on the security and privacy preservation of GANs and VAEs. Our focus is on the inner connection between attacks and model architectures and, more specifically, on five components of deep generative models: the training data, the latent code, the generators/decoders of GANs/VAEs, the discriminators/encoders of GANs/VAEs, and the generated data. For each model, component and attack, we review the current research progress and identify the key challenges. The paper concludes with a discussion of possible future attacks and research directions in the field.
\end{abstract}

}

\maketitle

\IEEEdisplaynontitleabstractindextext

%
\IEEEpeerreviewmaketitle

\ifCLASSOPTIONcompsoc
\IEEEraisesectionheading{\section{Introduction}\label{sec:introduction}}
\else
\section{Introduction}
\label{sec:introduction}
\fi

%
%
%
%

%
%

Over the past few years, computation power has advanced sufficiently to enable the success of deep neural networks in various applications.  Within this category, there are two categories of deep learning models: generative and discriminative. Generative models synthesize data we can observe in our world, such as plausible realistic photographs of human faces \cite{karrasProgressiveGrowingGANs2018}. Collectively, these are known as  deep generative models (DGMs). The other one is to divide observed data into different classes, e.g., face recognition, recommender systems, etc. \cite{masiDeepFaceRecognition2018}. This category of models is known as deep discriminative models (DDMs) \cite{jebaraMachineLearningDiscriminative2004}.

\begin{figure}
 	\centering
 	\setlength{\abovecaptionskip}{0.cm}
 	\includegraphics[width=0.5\textwidth]{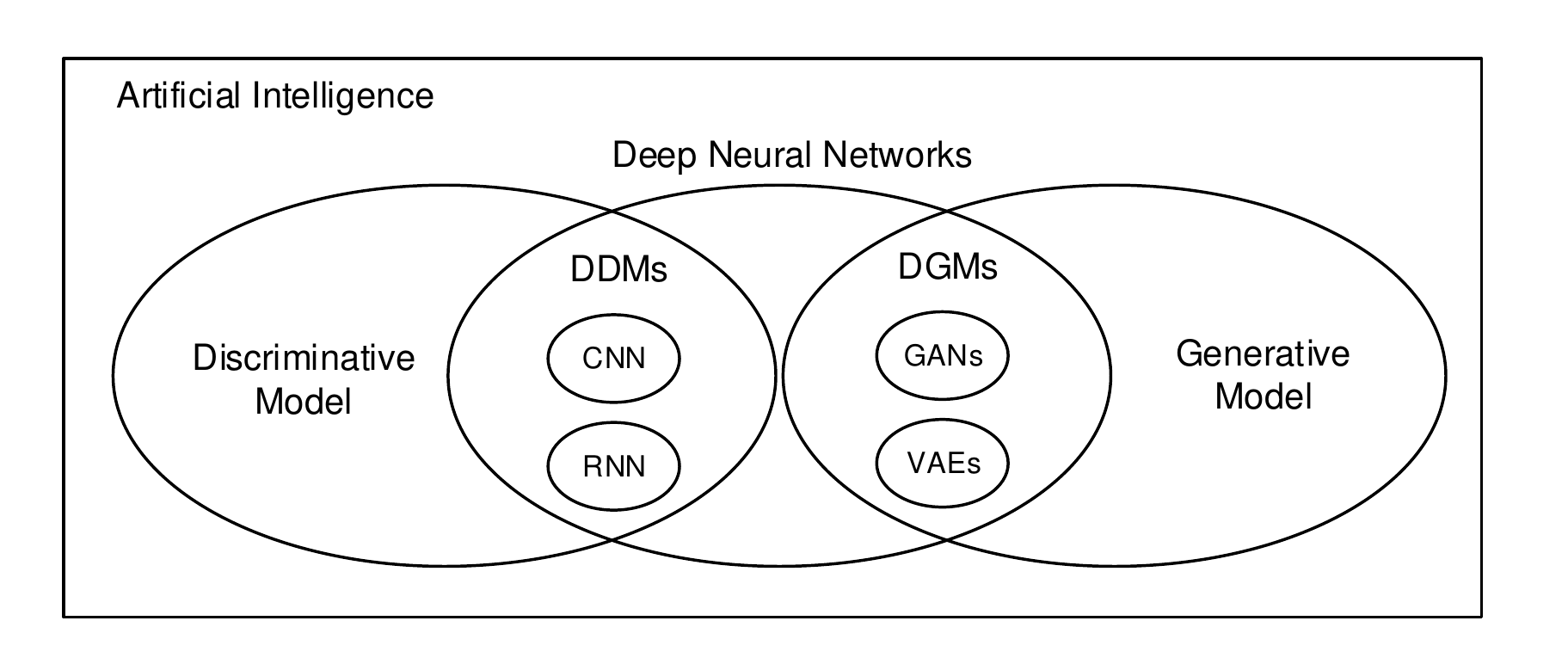}
 	\caption{The AI landscape. AI has two main branches, generative models and discriminative models. The deep neural network variants of these models have evolved into VAEs and GANs on the generative side and into RNNs and CNNs on the discriminative side.}
 	\label{relationship_diagram}
 \end{figure}

The most popular DGMs are generative adversarial networks (GANs) \cite{goodfellowGenerativeAdversarialNets2014} and variational auto-encoders (VAEs) \cite{kingmaAutoEncodingVariationalBayes2013}. Both are widely used to generate realistic photographs \cite{brockLargeScaleGAN2019}, synthesize videos \cite{vondrickGeneratingVideosScene2016}, translate one image into another \cite{isolaImagetoImageTranslationConditional2018}, etc. As the traditional DDMs, recurrent neural network (RNN) \cite{DBLP:conf/interspeech/MikolovKBCK10}, convolutional neural networks (CNN) \cite{GU2018354}, and their variants perform well at sentiment analysis \cite{medhatSentimentAnalysisAlgorithms2014}, image recognition \cite{heDeepResidualLearning2016}, natural language progressing (NLP) \cite{dengDeepLearningNatural2018,hirschbergAdvancesNaturalLanguage2015} and so on. A relationship diagram of the AI landscape is presented in Fig. \ref{relationship_diagram}.

As with any technology of wide influence, model security and privacy issues are inevitable. Naturally, any adversary will have two aspirations. The first is to sabotage the model so it does unsatisfactory work. The second is to breach privacy.  In sabotaging a model, for example, an attacker might turn a model that is supposed to generate human portraits into one that generates pictures of shoes \cite{pasquiniAdversarialOutdomainExamples2019}, or instead of correctly classifying pictures as pandas, it might classify them as gibbons \cite{goodfellowExplainingHarnessingAdversarial2015a}.  Breaching privacy might include stealing the training data or the whole trained model.  A famous example of this was where adversaries duplicated models trained by Amazon through black-box queries from APIs provided by its machine-learning-as-a-service  platform \cite{tramerStealingMachineLearning}. The same tactic has been used to restore the training set so as to acquire private information \cite{shokriMembershipInferenceAttacks2017a,fredriksonModelInversionAttacks2015}.

Poisoning attacks \cite{barrenoSecurityMachineLearning2010, papernotScienceSecurityPrivacy2016} and evasion attacks \cite{biggioEvasionAttacksMachine2013, goodfellowExplainingHarnessingAdversarial2015a} both attempt to force a model to do unsatisfactory work.  Poisoning attacks operate during the training phase, and attempt to compromise the model's abilities at the formation stage. Evasion attacks work during the test phase, with the aim of providing adversarial input to the trained model so that it produces unsatisfactory output. Adversarial input is generally called an adversarial example. 

At the component level, there are several different types of attacks. At the data level, we have membership inference attacks, which attempt to infer whether a given sample belongs to the model's training set \cite{shokriMembershipInferenceAttacks2017a} and, also, model inversion attacks, which try to reconstruct some or all of the training data based on the some prior information and the model's output \cite{fredriksonModelInversionAttacks2015}. At the attribute level, we have attribute inference attacks, which attempt to infer the sensitive attributes of data \cite{jiaAttriInferInferringUser2017}. Model extraction attacks work at the model level. These are a severe threat that try to duplicate the entire trained model \cite{tramerStealingMachineLearning}.

Although research into these attacks on DGMs is still in its infancy, there is a healthy body of literature on the security and privacy issues associated with DDMs. For instance, Papernot et al. presented a detailed adversarial framework of security and privacy attacks that included adversarial examples, strategies for membership inference attacks, and some defense methods \cite{papernotSoKSecurityPrivacy2018}. Focusing on scenarios and applications, Liu et al. categorized both the types of attacks and types of protection schemes \cite{liuWhenMachineLearning2021}. Serban et al. elaborated on adversarial examples, including their construction, defense strategies, and transfer capabilities \cite{serbanAdversarialExamplesObject2020}. As differential privacy is one of the most effective measure for mitigating privacy breaches, Gong et al. published a comprehensive review on differentially-private machine learning \cite{gongSurveyDifferentiallyPrivate2020}. A number of surveys have also been conducted on DDMs, particularly CNNs and RNNs, see e.g., \cite{litjensSurveyDeepLearning2017, liuSurveyDeepNeural2017, zhangSurveyDeepLearning2018,kamilarisDeepLearningAgriculture2018, zhangDeepLearningGraphs2020}.

In terms of DGMs, there has been much less work, as this survey will show. Our review unearthed the following research papers on: poisoning attacks \cite{dingPoisoningAttackDeep2019, salemBAAANBackdoorAttacks2020}; evasion attacks \cite{gondim-ribeiroAdversarialAttacksVariational2018, kosAdversarialExamplesGenerative2018, tabacofAdversarialImagesVariational2016, pasquiniAdversarialOutdomainExamples2019, yehDisruptingImageTranslationBasedDeepFake2020a, sunTypeAttackGenerative2020a, yangUASGUniversalMethod2020}; membership inference attacks \cite{mendelevitchFidelityPrivacySynthetic2021, chenGANLeaksTaxonomyMembership2019, hayesLOGANMembershipInference2019, hilprechtMonteCarloReconstruction2019, liuPerformingComembershipAttacks2019, stadlerSyntheticDataPrivacy2020a}; and attribute inference attacks \cite{stadlerSyntheticDataPrivacy2020a} and model extraction attacks \cite{huModelExtractionDefenses2021a}. To the best of our knowledge, there are no surveys devoted to the security and privacy of DGMs. However, in recent years, GANs and VAEs have advanced greatly so that, now, DGMs are attracting much more attention, both well-meaning and ill-intentioned. We therefore think it is time for a thorough survey of those attacks and, of course, their defenses. By comparing DGM attacks with DDM attacks and their known defenses, we may be able to identify some critical gaps between them.  

\begin{itemize}
	\item[$\bullet$] On a basic level, adversarial attacks are about the evolution of a  strategy. The attacks mentioned above were originally designed for discriminative models and DGMs have a very different purpose to DDMs. As such, the training algorithms and model architectures are also very different. Therefore, to perform traditional attacks against DGMs, the attack strategies must be updated. One single attack strategy cannot reveal the overall direction this evolution will take. Rather, a comprehensive review is required.
	\item[$\bullet$] Whether the evolved attacks will be general to DGMs is another concern. Since there are multiple variants of VAEs and GANs, such as beta-VAEs \cite{higginsBetaVAELearningBasic2016} and Wasserstein GANs \cite{arjovskyWassersteinGenerativeAdversarial2017} as well as other less popular types of DGMS, generality would make sense.
	\item[$\bullet$] There may be rare defense strategies specially designed for the occasions when a DGM suffers various types of attacks.
\end{itemize}

In these regards, a systematic study of the current state-of-play in the field will be essential to future research efforts. Thus, the main contributions of this survey include:
\begin{itemize}
	\item[$\bullet$] A brief introduction to VAEs and GANs, the most popular DGMs, beginning with their standard model structures and training procedures and ending with a comparison between the DGM and DDM architectures.
	\item[$\bullet$] An analysis of the feasibility of the various attacks given the two stated adversarial goals - to sabotage the model's proper functioning and to compromise privacy - and the vulnerability of the model's individual components. This section also categorizes the common attack strategies.
	\item[$\bullet$] A summary of the existing defense schemes and a discussion on the possible defense methods, which, given the rarity of defenses, make up the bulk of future research directions. 
	\item[$\bullet$] Suggestions for other fruitful research opportunities worthy of further attention.
\end{itemize}


%
%

\section{A Background on DGMs}



\subsection{Notations}

Consider a DGM with a training set $D_{train}$ that consists of numerous instances sampled from a real data distribution $P_{real}$ and an expectation that the training data distribution $P_{train}$ approximates the real data distribution $P_{real}$. The model learns the real data distribution from the training set and aims to generate samples that seem to be real but are unseen. Here, $x$ denotes a real sample in training set, $\hat x$ denotes a generated sample, and $D_{generated}$ and $P_{generated}$ denote the collection and distribution of the generated samples, respectively. For a generated data distribution $P_{generated}$ to be plausible, it must be close to the training data distribution, and therefore, in turn, close to the real data distribution. This can be expressed as $P_{generated}\approx P_{train} \approx P_{real}$. To maintain diversity, latent code $z$ is randomly sampled from a distribution defined as $P_z$. This is another representation of an input sample.

Both GANs and VAEs have two components, each taking the form of a neural network. A GAN consist of a generator and a discriminator; the corresponding functions are expressed as $f_{gen}$ and $f_{dis}$. A VAE consist of an encoder and a decoder with the corresponding functions similarly expressed as $f_{enc}$ and $f_{dec}$. Further, most evasion attacks involve a target output, denoted as $x_{target}$, and most membership inference attacks involve a query/series of queries $x_{query}$ the adversary uses to infer information. So, for instance, a membership inference might be explained as $Pr(x_{query}\in D_{train})$, where $Pr()$ denotes the possibility rate.

\begin{table}[htbp]
\renewcommand{\arraystretch}{1.3}
	\centering
	\caption{Notation}
	\begin{tabular}{|p{2cm}|p{6cm}|}
		\hline
		\makecell[l]{Notation} & \makecell[c]{Description} \\
		\hline
		$P_{real}$& Distribution of real data  \\
		\hline
		$D_{train}$, $P_{train}$ & Training dataset and its distribution\\
		\hline
		$x$    & Real sample \\
		\hline
		$\hat x$ & Generated sample \\
		\hline
		$D_{generated}$, $P_{generated}$ & Dataset of generated samples and corresponding data distribution\\
		\hline
		$z$ & Latent code\\
		\hline
		$P_z$& The latent distribution, such as Gaussian distribution\\
		\hline
		$x_{query}$& The given query sample for membership inference\\
		\hline
		$x_{target}$& The given target output sample for evasion attack\\
		\hline
		$f_{enc}$, $f_{dec}$ & Function of encoder and decoder\\
		\hline
		$f_{gen}$, $f_{dis}$ & Function of generator and discriminator\\
		\hline
		${KL}(\cdot || \cdot)$ & The Kullback--Leibler divergence\\ 
		\hline	
	\end{tabular}%
	\label{tab:notions}%
\end{table}%


\subsection{DGMs: GANs and VAEs}

As a major branch of deep learning, DGMs focus on data generation. DGMs are unsupervised, automatically learning the data patterns in a training set so that the model has the capacity to generate new samples in accordance with a distribution that is as similar as possible to the true data distribution. GANs learn the distribution implicitly under a minimax game where a generator tries to fool a discriminator and the discriminator tries not to be deceived \cite{oussidiDeepGenerativeModels2018}. VAEs learn the distribution explicitly by limiting reconstruction errors under an encoder-decoder framework. 

\begin{figure}
	\centering
	\setlength{\abovecaptionskip}{0.cm}
	\includegraphics[width=0.5\textwidth]{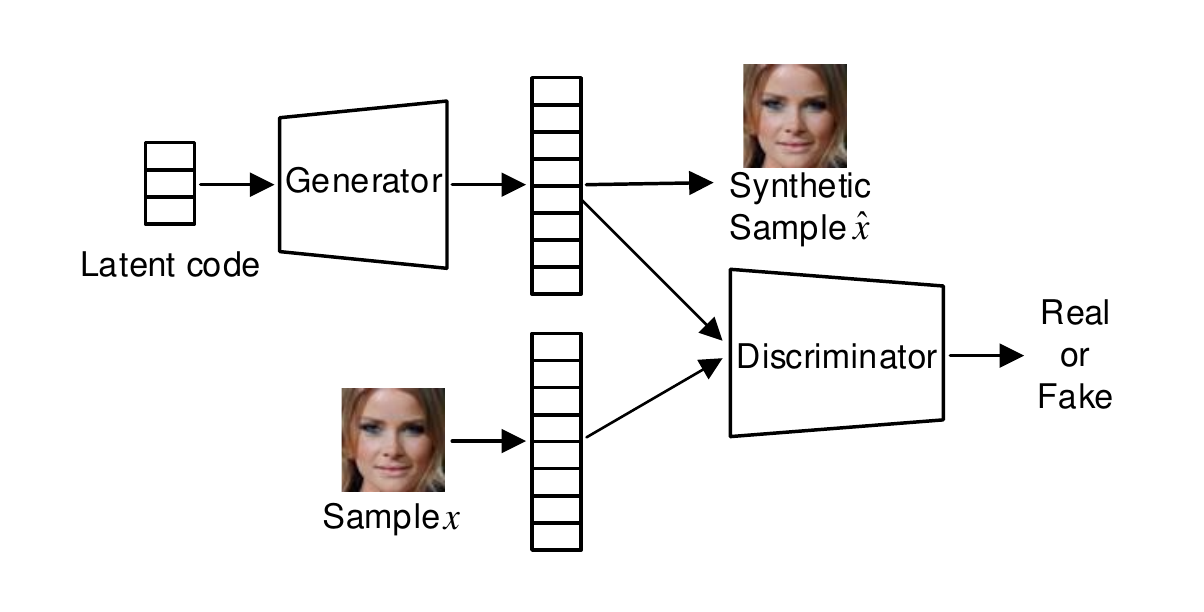}
	\caption{GAN architecture. A GAN consists of a generator and a discriminator, both of which are deep neural networks. In this example, the generator maps the latent code as an image representation, and the discriminator tries to distinguish between the generated image representation and the raw image representation. This ensures the generated samples are plausible. The process is formulated as a minimax game where generator tries to fool discriminator into classifying generated samples as raw samples and the discriminator tries not to be deceived.}
	\label{GAN_structure}
\end{figure}

\begin{figure}
	\centering
	\setlength{\abovecaptionskip}{0.cm}
	\includegraphics[width=0.5\textwidth]{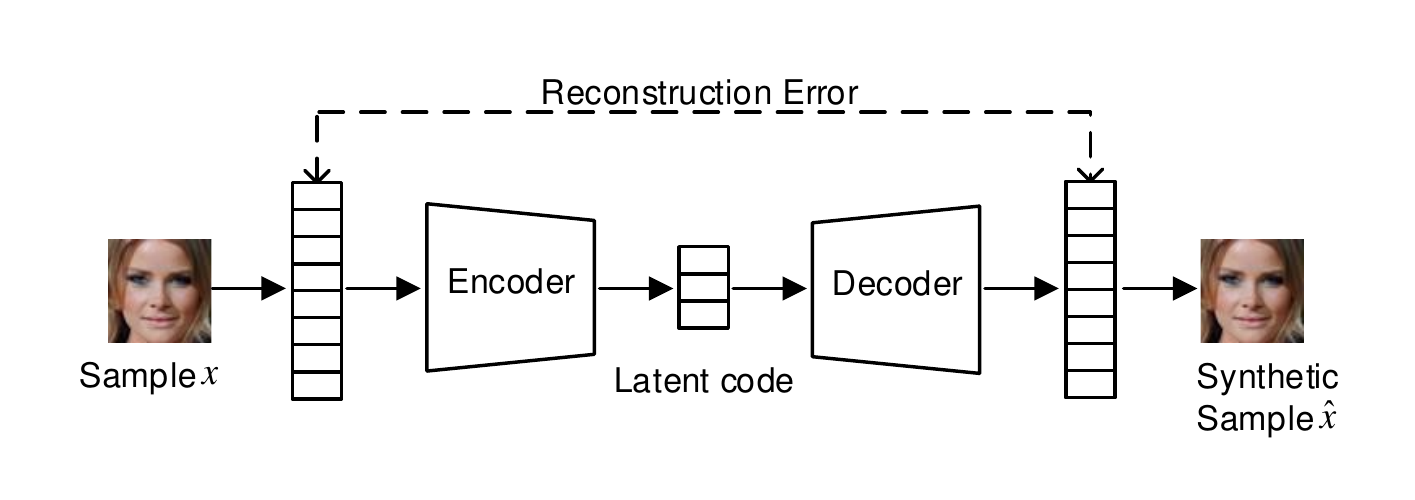}
	\caption{VAE architecture. A VAE consists of an encoder and a decoder, both of which are also deep neural networks. The encoder encodes the input representation into a lower-dimensional latent code, which subsequently is decoded into a representation by decoder. The reconstruction error mechanism between the input and decoded representations ensures the plausibility of the generated samples.}
	\label{VAE_structure}
\end{figure}

GANs and VAEs follow different principles and, thus, have different model architectures, as shown in Figs. \ref{GAN_structure} and \ref{VAE_structure}. However, both are made up of the same five broad components. These are:
\begin{enumerate}
    \item The training set, which consists of numerous real samples following a distribution that approximates the real data distribution. 
    \item Latent code, which is an alternative vector representation of the data. Typically, this has lower dimensionality than the input representation and is generally randomly sampled from a latent distribution to satisfy the requirement of never-seen generated samples. Essentially, the distribution is defined as latent space.
    \item Generator (GANs)/decoder (VAEs) - both are generative components that finish the mapping from randomly sampled latent code to a sample formally denoted as $z \rightarrow x$.
    \item Discriminator (GANs)/encoder (VAEs) - both are auxiliary components that help the generator/decoder become better trained and can, thus, be discarded when the training ends.
    \item Generated data, which is the output of the generator/decoder. With the well trained generator/decoder, the distribution of generated data will approximate the real distribution.
\end{enumerate}

\subsubsection{GANs} 
In GANs, the generator takes latent code as input and generates samples. Both these samples generated and real samples are then sent to the discriminator, which acts as a binary classifier with the task of distinguishing the real data from the generated data. Thus, a GAN's training is formulated as a minimax game \cite{myerson2013game} where a discriminator and a generator compete against each other. The generator tries to produce a fake sample that fools the discriminator into classifying it as true, while the discriminator tries to perfectly discriminate between the fake data and the true data. Formally, this can be expressed as
\begin{equation}
    \begin{aligned}
        \underset{f_{gen}}{min} \  \underset{f_{dis}}{max} \  L_{GAN} &=  E_{x\sim D_{train}}[log f_{dis}(x)] \\
        &+ E_{z\sim P_z}[log(1-f_{dis}(f_{gen}(z)))],
    \end{aligned}
    \label{GAN_loss}
\end{equation}
where $f_{gen}$ and $f_{dis}$ denote the generator and discriminator functions, respectively, $D_{train}$ denotes the training set, and $P_z$ is the prior latent distribution, usually a normal Gaussian distribution. 
The first term, denotes the real loss, i.e., the cross-entropy loss of the real data that is classified as real by the discriminator. The second term, denotes the fake loss, i.e., the cross-entropy loss of the generated data that the discriminator classifies as generated. The generator hopes to minimize the fake loss, while the discriminator hopes to maximize both the real and fake loss.

During training, the generator and discriminator are each trained in turn. While the discriminator is being trained, the parameters of generator network are fixed, and vice versa. The training ends when both the generator and discriminator are not showing further improvement. As a result, the generated data is so similar to real data that it successfully fools the discriminator.

\subsubsection{VAEs}  

VAEs generate samples based on the cascaded work of the encoder and decoder. The encoder compresses the input sample into a lower-dimensional latent space and the decoder decompresses randomly sampled latent code from the latent space into a sample. This compressing and decompressing is commonly referred to as encoding and decoding/reconstruction. A reconstruction mechanism is conducted between the input sample and decompressed sample so that the latent code keeps the maximal information of input sample during encoding process and the decompressed sample has minimal reconstruction errors during decoding. \textcolor{black}{As part of this process, the latent space must be regularized to be continuous and complete. Hence, a point randomly sampled from the latent space could be decoded as a new and plausible sample. The best encoding-decoding schemes and regularized latent spaces are achieved using an iterative optimization process with the loss function:}
\begin{equation}
    \begin{aligned}
    \color{black}
        min \ L_{VAE} =  -E_{z \sim Q(z|x)}[log P(x|z)] + KL(Q(z|x)||P(z)),
    \end{aligned}
    \label{VAE_loss}
\end{equation}
where $Q(z|x)$ and $P(x|z)$ are the encoder and decoder networks, respectively, $x$ is the input sample, and $z$ is the latent code. The first term, in the case of a reconstruction error, denotes the cross-entropy between the input $x$ and their reconstruction $\hat x$, $\hat x = f_{dec}(f_{enc}(x))$. The second term, often called regularization term, regularizes the latent space by ensuring the returned distribution $Q(z|x)$ is close to prior distribution of the latent code $P(z)$. Generally, this follows a standard multivariate Gaussian distribution $N(0, I)$. The Kullback-Leibler divergence $KL(\cdot||\cdot)$ is normally used to measure the distance between the two distributions.

\subsection{A Comparison of DGMs and DDMs}

\begin{table*}[htbp]
\renewcommand{\arraystretch}{1.3}
    \centering
    \caption{DGMs vs. DDMs}
	\begin{tabular}{|l|c|l|c|c|}
		\hline
		\multicolumn{1}{|l|}{}           & \multicolumn{2}{c|}{Model instance}                            & \makecell[c]{Input}                                                    & \makecell[c]{Model output}                 \\ \hline
		\multirow{4}{*}{DGMs} & \multirow{2}{*}{GANs}       & Generator          & Latent code                                              & Generated data               \\ \cline{3-5} 
		&                             & Discriminator      & Generated data, training data                            & A label[0,1] and a confidence score \\ \cline{2-5} 
		& \multirow{2}{*}{VAEs}       & Encoder            & Training data                                              & Latent code               \\ \cline{3-5} 
		&                             & Decoder            & Latent code                                            & Generated data          \\ \hline
		\multicolumn{1}{|l|}{DDMs} & \multicolumn{2}{c|}{CNNs} & Training data pairs \textless{}data, label\textgreater{} & Labels and corresponding confidence scores   \\ \hline
	\end{tabular}
	\label{dgms_comparison}
\end{table*}

To analyze DGMs' vulnerability to the mainstream attacks, e.g., membership inference attack, we compare DGMs with DDMs in aspect of components and corresponding inputs and outputs. Table \ref{dgms_comparison} lists the comparison result of the typical DGMs, i.e., GANs and VAEs, and a DDM, i.e., convolutional neural networks (CNNs). DDMs take the labels of each record in training set as input, which becomes the benchmark for training. For DGMs, from a holistic perspective, the real data is the concrete benchmark that verifies the quality of generated samples. And DGMs output data, while DDMs output the probabilities of a label, i.e., a confidence score. 

In view of the differences, there are clear security and privacy vulnerabilities for DGMs as follows.

\begin{enumerate}
\color{black}
    \item DGMs have more complex input, i.e., latent code and training data, which provides new directions for attacks against model input, like evasion attacks in latent space.
    
    \item DGMs reveal training data patterns by generating plausible samples, which leaves the privacy of training set rather transparent. Hence, DGMs are particularly vulnerable to attacks against training set, like membership inference attacks.

     \item DGMs, except discriminators of GANs, does not provide labels or confidence score, thus the derived attacks, e.g., membership inference attack in \cite{shokriMembershipInferenceAttacks2017a} and model extraction attack in \cite{tramerStealingMachineLearning}, are not perfectly feasible. 
\end{enumerate}

DGMs also have inner characters in common with DDMs. Both are based on deep neural networks, which means that DGMs tend to suffer from some of the same problems as DDMs - overfitting, for example. And discriminators of GANs are tantamount to DDMs. Further, both are trained on training data thus vulnerable to poisoning attacks. 

In general, DGMs are vulnerable to mainstream attacks, i.e., membership inference attacks, attribute inference attacks, model extraction attacks, poisoning attacks and evasion attacks; however, the traditional attacks, e.g., membership inference and model extraction attacks based on confidence score\cite{shokriMembershipInferenceAttacks2017a, tramerStealingMachineLearning}, would not work perfectly for DGMs. Specialized attack strategies are in requirement for attacking DGMs.

\section{Threat Models for Attacking GANs and VAEs}

\subsection{Adversary's Goals}

\textbf{Goal 1: Breaking the model}

Achieving this goal requires a disruption to the generative process that either results in: a) the intended output samples but at a low quality; b) some other presupposed samples within or out of the domain; or c) samples with no suppositions but ones that are not similar to the original output. If these presupposed samples sit within the domain, they should have a high probability of lying in the expected data distribution. Samples outside the domain would have the opposite and are not likely to follow the distribution.

This goal can be in achieved several ways. For instance, a poisoning attack can disrupt the model's generative abilities during training phase, while an evasion attack can do the same during the testing phase.

\begin{table*}[htbp]
\renewcommand{\arraystretch}{1.3}
    \caption{Adversarial Goals and Targets}
    \label{Adversarial Goals and Targets}
	\centering
	\begin{tabular}{|c|c|c|c|c|}
		\hline
		Adversarial   Goal                                               & Attack Target       & Attack Category                                                                                                               & Phase & Model Corruption \\ \hline
		\multirow{4}{*}{\shortstack{Goal 1: To break the model}} & Generator    & Evasion                                                                                                                       & Test           & Integrity                 \\ \cline{2-5} 
		& Decode          & Evasion                                                                                                                       & Test           & Integrity                 \\ \cline{2-5} 
		& Encoder                      & Evasion                                                                                                                         & Test           & Integrity                 \\ \cline{2-5} 
		& Dataset                      & Poisoning                                                                                                                       & Training       & Integrity                 \\ \hline
		\multirow{5}{*}{Goal 2: To steal confidential information}                           & Generator &\makecell[c]{ Membership   inference\\      Attribute inference\\      Model extraction} & Test           & Confidentiality           \\ \cline{2-5} 
		& Decoder    & \makecell[c]{Membership   inference\\      Attribute inference \\      Model extraction} & Test           & Confidentiality           \\ \cline{2-5} 
		& Discriminator                & Membership   inference                                                                                                          & Test           & Confidentiality           \\ \hline
	\end{tabular}
\end{table*}

\emph{Poisoning Attack}: The basis of this attack is to inject carefully crafted samples into the training set thereby poisoning it. Then, any model trained on the poisoned set will learn wrong abilities with wrong model parameters. Another way is to damage part of the model's structure, such as its loss function, to alter the model's workflow. Both strategies can be teamed with triggers to allow the attack to work within certain conditions, known as a backdoor poisoning attack \cite{chenTargetedBackdoorAttacks2017}. Such attacks tend to avoid early detection.

\emph{Evasion Attack}: This attack carefully crafts the model input to induce an unsatisfactory output. Such input is defined as an adversarial example. For a DGM, model input includes the latent code and the input sample. Accordingly, adversarial examples can be crafted for the latent code and sample, often called the latent adversarial example and adversarial example in this survey.


\textbf{Goal 2: Stealing confidential information}

Any information that authorized users could not obtain from a normal query to the trained model is confidential. Adversaries' prime targets include the model's parameters and its training set. Typical attack strategies include: the inference attack, where adversaries try to infer real data and/or attribute values with high confidence \cite{liProtectingIndividualInformation2007}; the model extraction attack, where the aim is to duplicate the functionality of the model \cite{tramerStealingMachineLearning}; and the model inversion attack, where adversaries recover the training data and thus also gain access to the model. To the best of our knowledge, there have not been any model inversion attacks against DGMs, yet. However, there have been several studies on how one might perpetrate a model inversion attack from the generated samples or latent code, while focus more on the latent space, such as the interpretability  \cite{shenInterpretingLatentSpace2020} and regularization \cite{patiAttributebasedRegularizationLatent2020a} of latent code. 

\emph{Membership Inference Attack} \cite{shokriMembershipInferenceAttacks2017a}: In this type of attack, the adversary tries to deduce whether a given sample is part of the model's training set. Prior information about the set, such as its size, can help them to deduce whether a set of samples are subset of the training set. With multiple queries, the entire training set might be recovered. Membership inference attacks lead to severe privacy leaks. They also provide clues about the strategies for other types of privacy attacks.

\emph{Attribute Inference Attack} \cite{stadlerSyntheticDataPrivacy2020a}: It is also known as record linkage attack, in the attack, adversaries have knowledge of some of the common attributes of the dataset, which they use to try and infer the sensitive attributes of a given sample. The common attributes are generally freely available to the public, such as a street scene, but the sensitive attributes are ones protected from public view, such as the number plates of the car parked along the street.

\emph{Model Extraction Attack} \cite{tramerStealingMachineLearning}: The idea of this attack is to infer the parameters or functions of the model via an efficient set of queries. If successful, the adversary can then copy the model's functions partly or even completely. 

\emph{Goal 1} is achieved by destroying the model's integrity and \emph{Goal 2} is achieved by destroying the model's confidentiality\cite{SCIAstrikesback2014}. Model integrity means that the model's training and testing process suffer no disturbance so the model produces normal output. Poisoning attacks disturb the training process and evasion attacks disturb the testing process, both of which result in unsatisfactory output, and accomplish \emph{Goal 1}. Model confidentiality means that sensitive data should only be disclosed to authorized users. Membership inference attacks and attribute inference attacks all reveal the training data, while model extraction attacks duplicate the functionality of the model. All of these attacks procure confidential information without authorization, thus accomplishing \emph{Goal 2}. A summary of these goals and targets is given in Table \ref{Adversarial Goals and Targets}.


\subsection{The Adversary's Prior Knowledge}

Most types of attack either rely on or work better when the adversary holds some prior information about the model or its training set. The more prior information the adversary holds, the more powerful the attacks and the more successful it is likely to be. Prior information that adversaries may have includes:

\begin{enumerate}
	\item \emph{Training data and training algorithm}. The security of the training data is the basis of the model's confidentiality. However, a model owner may publicly share their training set during testing to explain their algorithm, which could reveal much about that data and the model's parameters.
	\item \emph{Model parameters}. These include the discriminator and generator of GANs, and the encoder and decoder of VAEs. The model owners may publish a full GAN/VAE online to show their product and encourage further updates. Additionally, they may publish part of the model, i.e., the discriminator of GANs providing a tool to test the effectiveness of their work. With the model parameters, the adversaries can design more detailed and personalized attacks, i.e., specially crafting adversarial examples and inferring data membership.
	\item \emph{Latent code}. As another representation of data, latent code plays a decisive role in data synthesis. There are two situations by which adversaries could come to have this knowledge. First, they may have direct control of latent code, in which case, they can alter its value to satisfy their goals. Second, they may know a latent distribution, which means they can alter the latent code indirectly.
	
	\item \emph{Generated data}. This is the most easy and basic information for an adversary to get. Generally, it is acquired by querying a DGM through its API. Adversaries can also be provided with a set of generated data by an unknowing user.
\end{enumerate}

Adversaries also have types depending on their capabilities and the information they possess. If they have access to the training algorithms and data, they can act more like insiders to fundamentally corrupt the model. These are known as internal attackers. Those with access to only the generated data are called black-box adversaries. If with comprehensive knowledge of the model parameters, they are called white-box adversaries. If with no access to the model's parameters but have access to more than the generated data, for example, the latent code, they are defined as partial black-box adversaries. Table \ref{Adversarial_capability} lists these classifications.

\begin{table*}[htbp]
\renewcommand{\arraystretch}{1.3}
	\centering
	\caption{Adversary's information and capability. There are five categories of prior information an adversary can hold: training data and algorithms, model parameters, latent code, generated data, and other auxiliary (publicly collected) information. The confidentiality of the information decreases from left to right. $!$ denotes essential access, $\#$ denotes access not required, $\bigcirc$ denotes access possibly required.}
	\label{Adversarial_capability}
	\begin{tabular}{|l|c|c|c|c|c|}
		\hline
		\multicolumn{1}{|l|}{Adversary}& \multicolumn{1}{c|}{Training data and algorithm} & \multicolumn{1}{c|}{ Model parameters} & \multicolumn{1}{c|}{Latent code} & \multicolumn{1}{c|}{Generated data} & \multicolumn{1}{c|}{Auxiliary information} \\
		\hline
		Internal attack & $!$     & $\#$     & $\#$      & $\#$      & $\#$  \\
		\hline
		White-box attack &  $\bigcirc$     & $!$     &    $!$   & $!$     &  $\bigcirc$ \\
		\hline
		Partial black-box attack &   $\bigcirc$    &  $\#$     &    $\bigcirc$   & $!$     &  $\bigcirc$\\
		\hline
		Black-box attack & $\#$     & $\#$      &$\#$      &$!$     & $\#$  \\
		\hline
	\end{tabular}%
\end{table*}%

To reach their goals, adversaries consider the prior knowledge they have and design an attack strategy accordingly. 
Based on adversarial information against each components, including training set, latent code, GAN generator/VAE decoder, GAN discriminator/VAE encoder and generated data. For the fact that the latent code is the input of generator of GAN (decoder of VAE) and generated data is the output, we classify the attacks against latent code and generated data into the attacks against generator/decoder. Specifically, we get following types of attacks: attack against generator/decoder of VAE, attack against discriminator/encoder, and attack against training set. 
Table \ref{summary_on_adversarial_strategy_on_components} summarizes the literature on the types of attacks, the components targeted, and the strategies used for each type of DGM plus the data. Attacks on models/components that are not feasible are indicated as n/a. Attacks that are unexplored are marked as TBD to reflect this gap in the literature.

\begin{table*}[htbp]
\renewcommand{\arraystretch}{1.3}
	\centering
	\caption{Overview of the attack types by component. This table summarizes the basic strategy for each attack as it pertains to each component of the model. Non-existent situations are denoted as not applicable (n/a). Possible but still unexploited situations are denoted as to be determined (TBD).}
	\begin{tabular}{|m{6em}|m{12em}|m{11em}|m{8em}|m{8em}|m{8.5em}|}
		\hline
		\diagbox[width=6em,trim=l]{Component}{Attack} & \makecell[c]{Evasion} & \makecell[c]{Membership inference} &\makecell[c]{Attribute inference}  & \makecell[c]{Model extraction} & \makecell[c]{Poisoning} \\
		\hline
		Generator & {Forces a generator to do the wrong work via latent adversarial code. The adversarial code should not be too far away from the prior distribution of the existing \cite{pasquiniAdversarialOutdomainExamples2019}} & {Infers whether a given sample belongs to the training set based on the generated data \cite{liuPerformingComembershipAttacks2019, chenGANLeaksTaxonomyMembership2019, hilprechtMonteCarloReconstruction2019, mendelevitchFidelityPrivacySynthetic2021, goncalvesGenerationEvaluationSynthetic2020, stadlerSyntheticDataPrivacy2020a}} & {Infers sensitive attributes based on the generated data \cite{stadlerSyntheticDataPrivacy2020a}, 2020} & { Duplicates a model based on the generated data \cite{huModelExtractionDefenses2021a}} & \makecell[c]{n/a} \\
		\hline
		Discriminator & \makecell[c]{n/a} & {Infers whether a given sample belongs to the training set based on the  discriminator's output and exact confidence scores \cite{hayesLOGANMembershipInference2019}} & \makecell[c]{TBD}
      &  \makecell[c]{TBD}     & \makecell[c]{n/a} \\
		\hline
		Decode & {Forces a generator to do the wrong work via latent adversarial code \cite{sunTypeAttackGenerative2020a}} & {Infers whether a given sample belongs to the target training set based on the generated data \cite{liuPerformingComembershipAttacks2019, chenGANLeaksTaxonomyMembership2019, hilprechtMonteCarloReconstruction2019, mendelevitchFidelityPrivacySynthetic2021, goncalvesGenerationEvaluationSynthetic2020, stadlerSyntheticDataPrivacy2020a}} & {Infers sensitive attributes based on the generated data \cite{stadlerSyntheticDataPrivacy2020a}} & { Duplicates the target model based on the generated data \cite{huModelExtractionDefenses2021a}} & \makecell[c]{n/a} \\
		\hline
		Encoder & {Forces a generator to do the wrong work by making the image input adversarial, which makes the code adversarial \cite{sunTypeAttackGenerative2020a, yehDisruptingImageTranslationBasedDeepFake2020a, tabacofAdversarialImagesVariational2016, kosAdversarialExamplesGenerative2018, gondim-ribeiroAdversarialAttacksVariational2018, yangUASGUniversalMethod2020}} & \makecell[c]{TBD}      & \makecell[c]{TBD}      &  \makecell[c]{TBD}     & \makecell[c]{n/a} \\
		\hline
		Dataset & \makecell[c]{n/a} & \makecell[c]{n/a} & \makecell[c]{n/a} & \makecell[c]{n/a} & Disturbs model training process, including its static datasets and data processing logics \cite{dingPoisoningAttackDeep2019, salemBAAANBackdoorAttacks2020} \\
		\hline
	\end{tabular}%
	\label{summary_on_adversarial_strategy_on_components}%
\end{table*}%

\section{GANs: Attacks Against Generators} \label{GAN_attack}

This section elaborates on attacks against generators, targeting the input, i.e., latent code, or the output, i.e., generated data.



\subsection{Evasion Attacks}
Manipulating latent code is essential for mounting an evasion attack. An evasion attack can be explained as finding a latent code in a pre-set latent distribution that the generator maps into an unsatisfactory sample. The final generated sample is dissimilar to the original output, but similar to the target sample. Meanwhile, the corresponding latent code should be close to the original latent code. Otherwise, it might be detected by defenders when they verify whether the input latent code belongs to the pre-set distribution. Consequently, the loss function of an evasion attack consists of two parts: the adversarial term $L_{adv}$ to ensure the attack effect, and the regularization term $L_{reg}$ to regularize the perturbation. A hyperparameter $\lambda$ aims to balance the two parts: 
\begin{equation}
    \begin{aligned}
        L_{evasion} = L_{adv}+ \lambda \cdot L_{reg},
    \end{aligned}
\end{equation}
\begin{equation}
    \begin{aligned}
        L_{adv}=\Delta(x_{target},f_{gen}(z)).
    \end{aligned}
\end{equation}

Pasquini et al. \cite{pasquiniAdversarialOutdomainExamples2019} was the first to explore the evasion attack against generator. They assumed that a defender would build a distribution hypothesis test to check whether the latent code belongs to the prior distribution before the code is sent to generator. To pass validation, i.e., to ensure that the latent code follows the prior latent distribution, even with updating multiple iterations, they restricted the moment of the latent code to be close to the moment of a random variable sampled from the latent distribution. Formally,
\begin{equation}
    \begin{aligned}
    \color{black}
        L_{reg}=\sum_{i=1}^{k} \omega_{i}\left\|\mu_{z_0}(i)-\tilde{\mu}_{z}(i)\right\|_{2}^{2},
    \end{aligned}
\end{equation}
\textcolor{black}{where $z_0$ is the latent code randomly sampled from the prior latent space $p_z$, $\mu_{z_0}(i)$ is the $i_{th}$ moment of $z_0$, and ${\widetilde \mu}_z(i)$ is the $i_{th}$ sample moment of the latent vector $z$. Here, $z$ is the iteration result of the original latent code $z_0$. The parameter $\omega_i$ is the weight assigned to the $i_{th}$ moment difference.}

The full attack process would proceed by the adversary first sampling some initial latent code from the prior latent distribution. Then, they would adjust that code with gradient descent by minimizing $L_{evasion}$.

We believe this approach could be extended to the conditional generator, where the defender could randomly choose a label and expect the generator to output a sample of that label. Here, the adversarial loss function would be:
\begin{equation}
    \begin{aligned}
        L_{adv}=\Delta(x_{target},f_{gen}(z,y)),
    \end{aligned}
\end{equation}
where $y$ is the randomly chosen label by the defender. During the optimization process, $y$ remains constant.

Overall, it is worth noting that, in a properly functioning model, the generated data is intrinsically similar to the training data, and model overfitting exacerbates this similarity. Hence, generated data can be treated as substitute for the training data. Adversaries can breach much privacy through generated data, which brings us to membership inference, attribute inference, and model extraction attacks.

\subsection{Membership Inference Attacks (MIAs)}
"Membership" in our survey means whether a sample belongs to the ML model's training set. Since the generated data distribution of a DGM approximates its training set, the problem of inferring membership can be converted into a problem of determining whether the query sample follows the generated data distribution. In this way, membership inference means determining whether the query sample is close to the generated sample. If so, it probably belongs to the training set. Attacks derived from this idea are known as distance-based MIAs. Attribute inheritance provides another idea that, if a query sample is used to train a model, the generated data will preserve certain attributes of the query sample. These attacks are termed as attribute-based MIAs. Each is detailed below. Additionally, we introduce co-MIAs to expand the attacking scenarios and possible attack calibration strategy.

\begin{table*}[htbp]
\renewcommand{\arraystretch}{1.3}
	\centering
	\caption{The types of membership inference attacks. According to the target component and controlled adversarial information, adversaries design various attack strategies for a query sample (single MIA) or a set of query samples (co-MIA). In each category, we present required adversarial information and basic idea, further discussing whether it is applicable into co-MIA.}
	\begin{tabular}{|m{6em}|m{6em}|m{9em}|m{14em}|m{12em}|m{4.94em}|}
		\hline
		\multicolumn{1}{|l|}{MIA type}&\makecell[c]{Target}& \makecell[c]{Adversarial information}& \makecell[c]{Single MIA}& \makecell[c]{Co-MIA} & \makecell[c]{Paper, Year} \\
		\hline
		\multirow{6}{*}{\parbox{1\linewidth}{\vspace{5cm} Reconstruction\\distance}}&\multirow{6}[12]{*}{\parbox{1\linewidth}{\vspace{5cm} \shortstack{Generator/\\ decoder}}} & Query to generator/decoder & Attains $k$ generated samples and find the one closest to the query sample. If the distance is within the threshold, the sample is part of the target training set. & Applicable if adversaries know the preset size or overall belonging. Repeat single MIA for each sample in the query set and decide at the end. Until then, there is no information to share across different samples.  & \cite{chenGANLeaksTaxonomyMembership2019} 2019 \\
		\cline{3-6}    
		 &  &\multirow{2}[4]{*}{\parbox{1\linewidth}{
		\begin{enumerate}[leftmargin=*]
					\item Query to generator/decoder
					\item Access to latent code
		\end{enumerate}}} & Adjusts the latent code until the output sample is within the query sample threshold. Generally, employ an approximate optimization method. & Applicable, but it is not clear how to share information across different instances & \cite{chenGANLeaksTaxonomyMembership2019} 2019 \\
		\cline{4-6}    \multicolumn{1}{|c|}{} & \multicolumn{1}{c|}{}&\multicolumn{1}{c|}{}  & Gets $k$ latent code from a derived distribution and calculate the average distance between the generated samples and the query sample. & Applicable, but it is not clear how to share information across different instances. & \cite{hilprechtMonteCarloReconstruction2019} 2019 \\
		\cline{3-6}    
		 & & \multirow{2}[4]{*}{\parbox{\linewidth}{
		\begin{enumerate}[leftmargin=*]
					\item Query to generator/decoder
					\item Access to latent code
					\item Access to internals of generator/decoder
		\end{enumerate}}} & Adjusts the latent code until the output sample is within the query sample threshold. Generally, use an advanced first-order optimization algorithm, such as L-BFGS optimization. & Applicable, but it is not clear how to share information across different instances. & \cite{chenGANLeaksTaxonomyMembership2019} 2019 \\
		\cline{4-6}    \multicolumn{1}{|c|}{}&\multicolumn{1}{c|}{} & \multicolumn{1}{c|}{} & Builds an adversary network like the encoder. Adjust its parameters until the output latent code is mapped as a sample close to the query sample & Applicable. Information can be shared across different instances.  & \cite{liuPerformingComembershipAttacks2019} 2019 \\
		\cline{3-6}    
		& &\parbox{1\linewidth}{\begin{enumerate}[leftmargin=*]
				\item Query to generator/decoder
				\item Partial real data
		\end{enumerate}}& Performs a simple membership inference test by randomly checking the similarity between the generated data and the training samples. & None  & \cite{mendelevitchFidelityPrivacySynthetic2021} 2021 \\
		\hline
		$\epsilon$-ball distance& \shortstack{Generator/\\ decoder} & Query to generator/decoder & Finds the number of generated samples in the $\epsilon$-ball neighborhood of the query sample. & Applicable, but it is not clear how to share information across different instances.  & \cite{hilprechtMonteCarloReconstruction2019} 2019 \\
		\hline
		Attribute &\shortstack{Generator/\\ decoder}& \parbox{\linewidth}{\begin{enumerate}[leftmargin=*]
				\item Query to generator/decoder
				\item Reference dataset
				\item Training algorithm
		\end{enumerate}} & \vspace{7pt}Checks whether the generated data has inherited the attributes of query sample & \vspace{7pt}Applicable & \cite{stadlerSyntheticDataPrivacy2020a} 2020 \\
		\hline
		Discriminator &Discriminator& Query to discriminator & The confidence score of the query sample is extremely high. & Applicable, but it is not clear how to share information across different instances  & \cite{hayesLOGANMembershipInference2019},2019 \\
		\hline
	\end{tabular}%
	\label{tab:addlabel}%
\end{table*}%

\subsubsection{Distance-based MIA}
We define the membership inference based on the distance between the query sample and the generated samples for two reasons. First, from the perspective of distribution approximation \cite{liuPerformingComembershipAttacks2019, chenGANLeaksTaxonomyMembership2019}, since the output distribution approximates the training data distribution, the probability that the query sample belongs to the training set is proportional to the probability that the query sample belongs to the output data distribution. Therefore, the inference can be expressed as whether one of the query samples belong to the output distribution - or, more specifically, whether the query sample was produced by the target generator. Second, from the perspective of overfitting the model \cite{hilprechtMonteCarloReconstruction2019}, if there are several generated samples close to the query sample, the query sample is probably a member of the training set. Both ideas revolve around whether one or more generated samples are close to the query sample. Formally, this can be expressed as 
\begin{equation}
    \begin{aligned}
        \underset{x \in G(\cdot)}{min} \Delta(x_{query}, x),
    \end{aligned}
\end{equation}
where $x_{query}$ is the given sample, $x$ is the generated sample from target generator $G$, and $\Delta$ is the distance function that calculates the distance between two samples. 

\begin{figure}
	\centering
	\setlength{\abovecaptionskip}{0.cm}
	\includegraphics[width=0.33\textwidth]{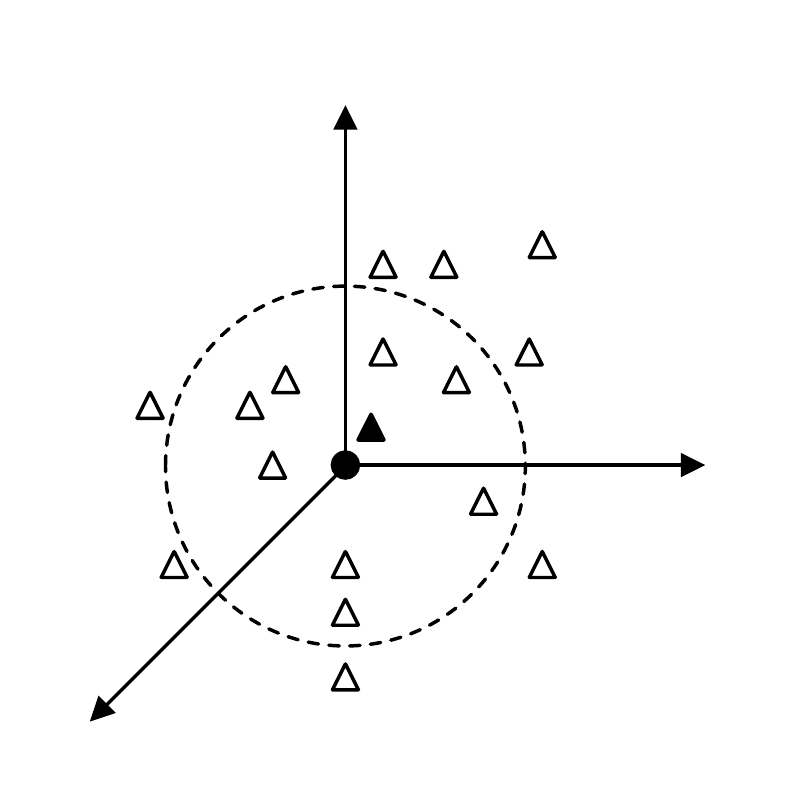}
	\caption{Membership inference attacks: reconstruction distance and $\epsilon$-ball distance. The dots ($\bullet$) denote the query sample, the triangles ($\triangle$ and $\blacktriangle$) denote the generated samples, and the dashed circle denotes the threshold of $\epsilon$-distance. With reconstruction distance, the sample that has the closet distance (labeled as $\blacktriangle$) is the focus. With $\epsilon$-ball distance, the quantity of generated sample in the dashed circle is the focus (quantity = 9).}
	\label{MIA_syn_data}
\end{figure}

In the papers we reviewed, most calculated these distances using either $\epsilon$-ball, where quantity was the focus, or reconstruction distance when quality mattered. Fig. \ref{MIA_syn_data} provides more details.

\textbf{Reconstruction Distance} Herein, adversaries focus on the quality of the distance measure. In other words, they are trying to find the generated sample that is closest to the query sample. The closest generated sample is called the reconstruction of the query sample, and the distance between the two is called the reconstruction distance - formally,
\begin{equation}
    \begin{aligned}
        R(x_{query},|G) = \underset{x \sim G}{argmin}  \Delta(x_{query},x)
    \end{aligned}
\end{equation}.
Empirically, it is impossible to obtain every generated sample and find the closest one. Multiple solutions have therefore been proposed to solve the optimization problem based on limited prior information. These are outlined as follows:

\begin{itemize}
	\item 
	Chen et al. considered a black-box attack and simply calculated the distance between the query sample and each generated sample \cite{chenGANLeaksTaxonomyMembership2019}. The sample with the least distance to the query sample was deemed the reconstruction. A judgment was then made about the reconstruction error.
	
	\item If the adversary has access to the latent code, they can adjust it to get an optimal solution in the regularized latent space. Chen et al. \cite{chenGANLeaksTaxonomyMembership2019} proposed to approximate the optimum via Powell's Conjugate Direction Method \cite{powellEfficientMethodFinding1964}, while Liu et al. \cite{liuPerformingComembershipAttacks2019} proposed building another set of neural networks to find the optimal latent code. These adversarial networks took the query sample as input and output the latent codes, like encoders. The adversary then adjusted the parameters of the adversarial net until the output latent code reaches the optimum. This approach essentially transforms the optimization problem into a parameter tuning exercise. However, without the generator's gradient information, Liu and colleagues used finite-difference to approximate the gradient and find the optimum latent code.
	
	\item White-box attackers, i.e., attackers with access to the internals of the generator, including the gradient information, can solve the optimization problem more accurately by using an advanced first-order optimization algorithm, such as L-BFGS \cite{chenGANLeaksTaxonomyMembership2019, liuPerformingComembershipAttacks2019}. Such a solution would be suitable for solving both optimization problems - for the latent code or the parameters.
\end{itemize}

Hilprecht et al. \cite{hilprechtMonteCarloReconstruction2019} made a compromise in cases where a precise reconstruction was not required to calculate the average distance between the query sample and each generated sample. 

Model publishers sometimes launch an MIA themselves before publishing the model to evaluate the model's security.This process is more commonly called a membership inference test, as shown in Fig. 5. Some researchers have proposed an easier test that is also based on the distance between the generated sample and the query sample \cite{mendelevitchFidelityPrivacySynthetic2021,goncalvesGenerationEvaluationSynthetic2020}. In the test, the ``adversaries'' have no knowledge of the model but full knowledge of the training data. The test works as follows:

\begin{enumerate}
	\item The raw data is randomly split into two disjoint subsets of equal size, $D=D_1 \cup D_2$. The generative model is trained on $D_1$ and a dataset of generated data $D_{generated}$ is produced.
	\item The adversary has access to a subset of $D$, denoted $D_3$. And samples in $D_3$ may belong to either $D_1$ and/ or $D_2$.
	\item Given a query sample $x$ from $D_3$ and the disclosed generated dataset $D_{generated}$, the adversary calculates the distance between the query sample and each sample in generated dataset with  $\Delta(x,\hat x)$, where $x \in D_3$ and $\hat x \in D_{generated}$.
	\item The adversary determines that $x$ is part of the training set $D_1$ when $\Delta(x,\hat x)$ is lower than some threshold. Value 2, 3 or 5 is recommended for threshold with a Hamming distance.
\end{enumerate}

If the test has a high success rate (above 0.5), the inference is better than a random guess and definitely effective. The model publisher might administer a membership inference test to several candidate models and choose the one with the least success rate. It is worth noting, however, that membership inference tests are meaningful for model publishers to validate the risk of disclosing the membership privacy but this does not necessarily translate into a practically secure model with the precondition that the adversary is capable of the training data.

\begin{figure}
	\centering
	\setlength{\abovecaptionskip}{0.cm}
	\includegraphics[width=0.5\textwidth]{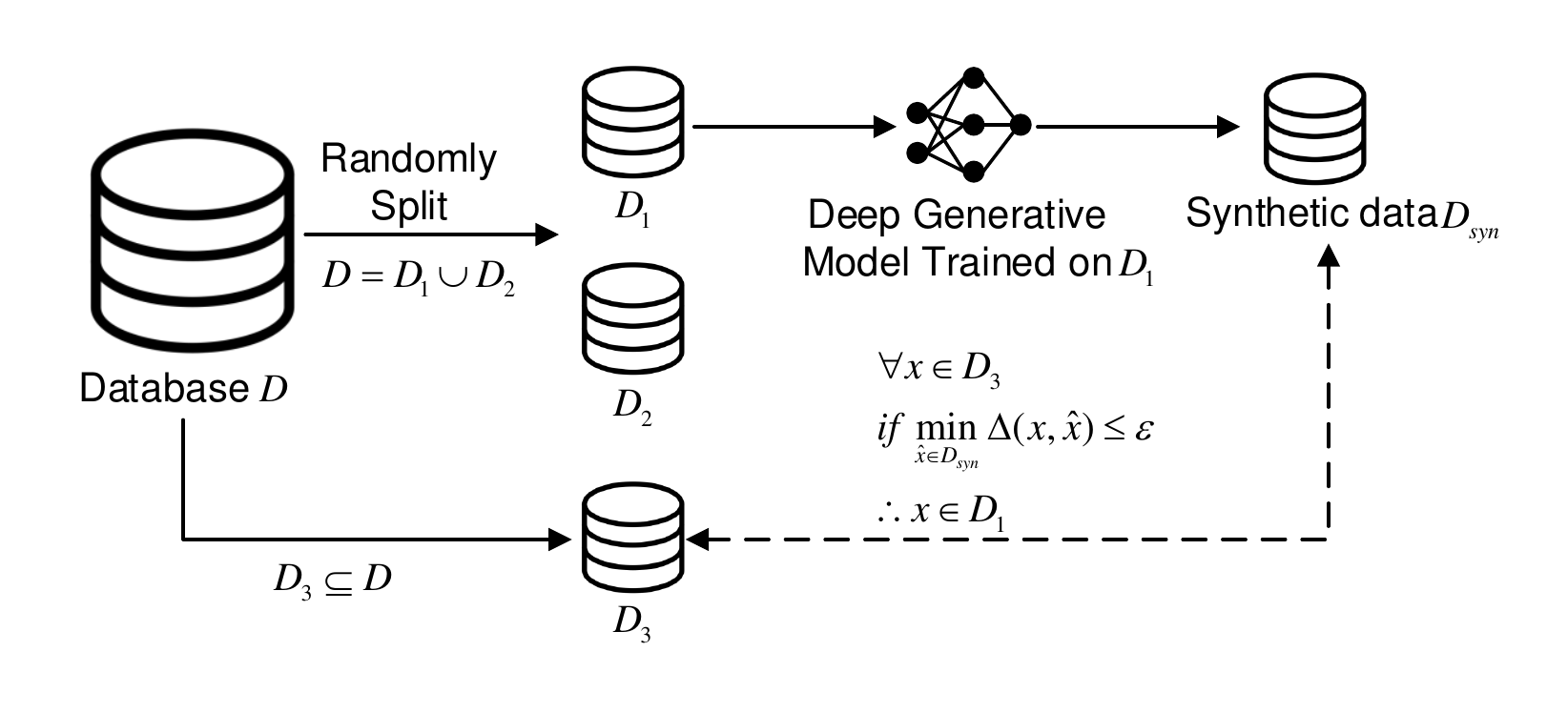}
	\caption{Membership inference test. The model publisher has full control of training data and test the confidential level of the candidate models. A single membership inference test is launched for each sample in $D_3$ and the publisher obtain the overall success rate. The higher the success rate, the less confidential information the candidate model has.}
	\label{MI_test}
\end{figure}

\textbf{$\epsilon$-ball Distance}. With $\epsilon$-ball distance, the adversary is more concerned with quantity than quality - the reason being that the more generated samples around the query sample, the more likely the sample is of the target training set. The attack is launched as:
\begin{enumerate}
	\item Define the $\epsilon$-neighborhood of the query sample as $U_{\epsilon}(x_{query}) = \Delta(x,x_{query})\le \epsilon$.
	\item Obtain a generated dataset by querying the generator or getting on directly from the model publisher.
	\item Calculate the distance between each generated sample and the query sample and count how many samples are in the $\epsilon$-neighborhood of the query sample $U_{\epsilon}(x_{query})$.
	\item If co-MIA is launched, calculate the average quantity and compare the two results.
\end{enumerate}

Hilprecht et al. \cite{hilprechtMonteCarloReconstruction2019} was the first to come up with this idea and, further, these authors initially tried to incorporate exact distances into Step 3. So, if the generated data was in the $\epsilon$-neighborhood of the query sample, they recorded the distance; if not, they ignored the sample. The alternative was to calculate the average distance between the query samples and the generated samples in the $\epsilon$-neighborhood of the query sample. However, the empirical results show that there were no significant differences between the basic two ideas. Therefore, the samples in the $\epsilon$-neighborhood were taken to play the main role in the attack.

Notably, an appropriate choice of $\epsilon$ is crucial for the success of this attack. Two heuristics are used, i.e., percentile and median, with the empirical results showing that the median heuristic outperforms the percentile. Interested readers can refer to \cite{hilprechtMonteCarloReconstruction2019} for more details.

\subsubsection{Attribute-based MIAs}
Attribute-based MIAs is based on the query sample's impact on the DGM's output distribution. To implement this attack, Stadler et al. \cite{stadlerSyntheticDataPrivacy2020a} propose shadow training, which requires prior knowledge of a reference dataset, the training algorithm, and a generated dataset from the target model. The reference dataset must follow the same distribution as training set, and the two datasets may overlap. The shadow training procedure works as follows:

\begin{enumerate}
	\item Make two kinds of shadow training sets, one containing the query sample and the other does not. Then randomly sample data from the reference dataset to form multiple data sets. Half should include query sample.
	\item Run the training algorithm on each shadow training set and collect the generated samples of the shadow model. If the shadow training set contains the query sample, the generated samples should be labeled with 1, otherwise 0. This results in data pairs $<generated \  data, 1 \  or \  0>$.
	\item Train a binary classifier on the data pairs.
	\item Use the trained classifier to predict the label with the generated data. If confidence score is above 0.5, the query sample belongs to the training set.
\end{enumerate}

To reduce the effect of high-dimensionality and sampling uncertainty, Stadler et al. suggest to use feature extraction techniques on the collected generated samples before training the classifier. In this way, the aim becomes detecting the target's influence on the feature vector, not on the whole image. Whether the attack is successful depends on two factors - first, whether the target's presence has a detectable impact on any of the features; second, whether the generated dataset has preserved these features from the raw data, i.e., preserved the target's signal.

\subsubsection{Co-MIAs}
When adversaries control certain additional information about the training set of target model, they tend to launch co-MIAs on a set of samples. Co-MIAs have evolved from the single MIA. Several co-MIA scenarios are possible depending on the additional information that the adversary holds, which are listed as follows.

\emph{The Preset Size}: In this situation, the adversary knows that $n$ samples in the query set belong to the training set. Hence, they launch a single MIA using each of those samples and sort the results by the degree from the distance function. The top $n$ samples are regarded as the training data. This type of co-MIA is an overlay of several single MIAs. Generally, either distance function is feasible. Hilprecht et al. \cite{hilprechtMonteCarloReconstruction2019} used the $\epsilon$-ball distance for GANs and reconstruction distance for VAEs. 

\emph{Overall Belonging}: In this situation, either all or none of the query samples belong to the target training set. The adversary still launches a single MIA against each sample, but then calculates the average based on the degree derived from the distance function. This average is the final determiner of whether all or none belong to the target training set. There are two ways to calculate the average. In the first way, each single sample is checked to ascertain whether it belongs to the training data and then adversaries count the samples that they believe to be in. If most of the query samples are in, then so is the whole set \cite{liuPerformingComembershipAttacks2019}. The second way is to calculate the average of the distance function's output and make a judgment based on that average. Additionally, if the adversary uses a neural network to finish the reconstruction, as mentioned in the single MIA, they can co-train one single model with all the query samples. Then the overall loss will be defined as the average of the loss for each sample \cite{liuPerformingComembershipAttacks2019}. If the adversaries mount an attribute-based MIA, they simply need to change Step 1 from the query sample to the query dataset, so that one shadow training set contains the query dataset and the other does not.

\subsubsection{Attack Calibration}
Chen et al. \cite{chenGANLeaksTaxonomyMembership2019} found it easier to generate a close sample for a less complicated query sample with an arbitrary generator. Likewise, it may be more difficult for a more complicated sample with the target generator. To mitigate the dependency on the representation of the query sample, they designed a reference generator which is trained on a relevant but disjoint dataset and mounted the same MIA against it, providing reference for final membership inference. In their research, they used the reconstruction distance as a main tool. As such, they deemed that if the reconstruction of the target DGM was close to the query sample, while the reconstruction of the reference DGM was far way, the query sample was more likely belonging to the training set.

\subsection{Attribute Inference Attacks}

Attribute inference attacks in this survey specifically target the attributes of data that should remain private. In this attack, an adversary tries to infer the private attributes of a data record based on other public attributes that are easily accessible. The most common source of these public attributes is the generated data. To be useful, this data's attributes must be plausible, but those plausible attributes simultaneously reveal patterns in the data. Adversaries can then look for inner connections between the private and common attributes. Once those connections become concrete, the private attributes will be revealed. The key to attribute inference attacks is therefore to find the inner connections between the data attributes.

Stadler et al. \cite{stadlerSyntheticDataPrivacy2020a} simplified this attack to a setting where there was only one sensitive attribute with a value in the continuous domain. They then formalized the problem as a regression problem in which an attacker learns to predict the value of an unknown sensitive attribute from a set of known attributes, with access to a dataset (either raw or generated).

In detail, the adversary is capable of a generated data set, denoted as $D$. The attributes of each sample are split into a disclosed part and a private one. The disclosed part contains all the attributes that the adversary can collect publicly, e.g., information on social networks, denoted $R_k = [r_1,r_2,\cdots,r_k]$, supposed to have $k$ disclosed attributes. The private part is what the adversary targets, denoted as $r_s$. The regression problem is solved by

\begin{equation}
    \begin{aligned}
        r_s=R_k \cdot w_{D}+\xi, \xi_i \sim N(0,\sigma ^2).
    \end{aligned}
\end{equation}
When the training ends, the adversary can derive the sensitive attribute of the target sample by calculating $r_{s}= w_{D} \cdot R_k$, where $R_k$ denotes the $k$ known attributes of the target sample. Intuitively, if the accessible set $D$ contains more training samples, the regression prediction can be more valid.


Furthermore, such method can verify whether the generated data mitigates the risk of attribute leakage with the bulk of training dataset involved \cite{stadlerSyntheticDataPrivacy2020a}. Two regression models are built on generated data set and raw training data set respectively. In this way, the private attribute of the query sample has two predictions. If the model based on the raw training dataset has more accurate predictions, the generated images definitely protect the attribute privacy, thus reducing the adversary's chance of success.

\subsection{Model Extraction Attacks}
The goal of the model extraction attack is to build a local model to clone the target model. Here, due to the approximation of generated and training data distribution, a straightforward approach is to train the local model on the generated data of the target model. The key to such a model extraction attack is to acquire generated data that are highly similar to the training data.

Based on the idea, Hu and Pang \cite{huModelExtractionDefenses2021a} collected generated samples to train a local DGM to approximate the target model. \textcolor{black}{They then designed two types of GAN attacks - accuracy extraction attack and fidelity extraction attack, targeting the model's data distribution, i.e., the generated data distribution of the target model, and the model's training set, i.e., the training data distribution of the target model.}

\subsubsection{Accuracy Extraction}

At this stage, the adversary trains a local GAN to extract the target model by minimizing the difference between the generated data distribution of the local model and that of the target. The extraction needs a large amount of generated data, otherwise the performance of local model is poor due to insufficient training samples. However, Hu and Pang found that more generated data did not always result in a better local model \cite{huModelExtractionDefenses2021a}. The more they queried, the more poor-quality samples they retrieved, which comprised the success of the attack. Therefore, adversaries have to control the quantity of generated data.

\subsubsection{Fidelity Extraction}
As with accuracy extraction, stealing the training data distribution of the target model is also formulated as a problem of minimizing the difference in distributions between the local generated sets and target training sets. To accomplish this, Hu and Pang propose two prior knowledge scenarios, noting that, either way, success requires at least some non-generated samples:

\begin{enumerate}
	\item Partial black-box fidelity extraction: generated data and some real samples from the training set; or
	\item White-box fidelity extraction: generated data, some real samples, and the discriminator of the target model.
\end{enumerate}

With the partial black-box version of the attack, the adversary retrains a local model on the generated data and continues training after adding in the available real data. In the paper, 50,000 generated samples were used. With the white-box attack, the adversary first leverages the discriminator to subsample the generated samples. Then, the local model is trained on the refined samples and further retrained on the available real data. Note that, considering some discriminators output a score rather than a probability, the discriminator was calibrated on real samples from the target GAN's training set through logistic regression. By comparison, white-box adversaries need to query both the generator and discriminator, and require more generated samples for subsampling.

\subsection{Summary}
As the generative component of GANs, generators often provide more information than imagination. They tend to be sensitive to trivial perturbations of latent code and, thus, can be vulnerable to evasion attacks. Additionally, because the generated data distribution approximates the training data distribution, generated samples reveal confidential information somewhat by design. This makes generators particularly vulnerable to MIAs, attribute inference attacks, and model extraction attacks.

\section{GANs: Attacks Against Discriminators}

For discriminators in GANs, they are deep binary classifiers that distinguish generated data from training data, which motivates the generators to produce more plausible samples. Since discriminator play no part in testing, and their output is not worth stealing, the only attack that applies to discriminators is the MIA. Even here, to the best of our knowledge, the only study on discriminator MIAs was published by Hayes et al. \cite{hayesLOGANMembershipInference2019}.

\subsection{MIAs Against Discriminators} \label{Whitebox_MIA}

As a deep binary classifier, if overfits, the discriminator would output extremely high confidence score for training samples and significantly low confidence score for generated samples. Hayes et al. first proposed MIAs on target discriminator \cite{hayesLOGANMembershipInference2019}. The attack strategy is simple: the adversary inputs the query sample into the target discriminator which subsequently outputs a confidence score. If the confidence score is above a threshold, e.g., 0.9, the query sample is part of the target training set with high possibility. Obviously, however, the attack does require direct access to the discriminator; hence, this is a white-box or internal attack.

\subsection{MIAs Against Shadow Discriminators}
As discriminators are not always accessible or even retained after training, there is a second and more complicated (partial) black-box version of MIAs that involves a shadow discriminator \cite{hayesLOGANMembershipInference2019}. This shadow discriminator is an approximate copy of the target that, once built, is targeted with the attack outlined in Section \ref{Whitebox_MIA}.

To build the shadow discriminator, the adversary collects samples that are in and out of training dataset of target model, separately defined as real and fake data. Auxiliary information includes: 1) samples generated by the target generative model. Given a well-trained generative model, the generated samples should be similar enough to fool the discriminator into regarding them as real data;  and 2) any additional information the adversary can collect, such as samples found online. The setting has a practical significance since most models are built from public data. The adversary then labels the collected samples with "real" or "fake" label to form a training set for the shadow discriminator.

If the adversaries can only collect fake data that were not used to train the target model, such as samples collected from online or testing set of target model, they can collect the generated samples as real data. Then the shadow discriminator is trained on fake and real set. If the adversaries can not collect fake data but limited real data, or even no auxiliary data, the adversary can collect the generated samples as real data, and build a local GAN to generate fake data. The local GAN is trained on the collected real data, and When the training ends, the local discriminator is regarded as the shadow discriminator. If the adversaries successfully attain a subset of real set and fake set, to train a local GAN or an alone discriminator is feasible. With a shadow discriminator in hand, the adversary can infer data membership with a white-box attack. Further, the target model could be any DGMs, not just GANs. 

\subsection{Co-MIAs}
Co-MIAs are also based on basic idea in Section \ref{Whitebox_MIA}. These attacks are designed to recover the target training set when the size of target training set known \cite{hayesLOGANMembershipInference2019}. Specifically, the adversary launches an MIA for $n+m$ query samples against a target or shadow discriminator, where $n$ is the size of the training set, and $m$ is the number of datapoints that do not belong to the training set. Then the discriminator outputs the confidence score. The adversary sorts the scores in descending order and the top $n$ samples are regarded as target training set. Table \ref{CoMIA_discriminator} summarizes two situations of Co-MIA against target or shadow discriminators.

\subsection{Summary}
Though not directly involved in data generation, the discriminators of GANs can reveal data membership with sophisticated adversaries. For the model security and privacy, it is essential to realize the importance of the discriminator and not to expose it.

\begin{table*}[htbp]
\renewcommand{\arraystretch}{1.3}
	\centering
	\caption{Summary of co-MIAs against discriminators. 
	}
	\resizebox{\textwidth}{!}{
		\begin{tabular}{|m{6em}|m{5em}|m{9em}|c|m{6em}|m{13em}|m{9em}|}
			\hline
			\diagbox[width=6em,trim=l]{Attack}{Category} & \multicolumn{1}{c|}{Target model} & \multicolumn{1}{c|}{Prior information} & \multicolumn{1}{c|}{Auxiliary dataset} & \multicolumn{1}{c|}{Local model} & \multicolumn{1}{c|}{Source of real data} & \multicolumn{1}{c|}{Source of fake data} \\
			\hline
			\makecell[l]{White-box} & \makecell[c]{GANs} & \multicolumn{1}{p{9em}|}{\parbox{\linewidth}{\begin{enumerate}[leftmargin=*]
					\item The size of the training set
					\item The discriminator of the target model
		\end{enumerate}}} & n/a & \makecell[c]{n/a} & \makecell[c]{n/a} & \multicolumn{1}{c|}{n/a} \\
			\hline
			\multirow{8}[8]{*}{Black-box} & \multirow{8}[8]{*}{\shortstack{Generative\\ models}} & \multicolumn{1}{l|}{\multirow{8}[8]{*}{\shortstack{The size of the training\\ set}}} & \multicolumn{1}{p{12em}|}{The samples that were not used to train the target model} &\makecell[c]{Discriminator} & \multirow{2}{*}{The target generated samples} & \multirow{2}{*}{Auxiliary dataset} \\
			\cline{4-7}          
			&       &       & \multicolumn{1}{p{12em}|}{The samples that were used to train the target model} & \makecell[c]{GAN} & \parbox{\linewidth}{\begin{enumerate}[leftmargin=*]
					\item The target generated samples
					\item Auxiliary dataset
		\end{enumerate}} & Samples generated by local generator \\
			\cline{4-7} 
			&       &       & \multicolumn{1}{l|}{No auxiliary dataset} & \makecell[c]{GAN} & \multicolumn{1}{l|}{The target generated samples} & Samples generated by local generator \\
			\cline{4-7}          &       &       & \multirow{3}[2]{*}{\makecell[l]{Both the training and test set\\ samples}} & \makecell[c]{Discriminator} & \parbox{\linewidth}{\begin{enumerate}[leftmargin=*]
					\item The target generated samples
					\item Auxiliary training set
		\end{enumerate}} & \multicolumn{1}{l|}{Auxiliary test set} \\
			\cline{5-7} &       
			&       &       & \makecell[c]{GAN} &\parbox{\linewidth}{\begin{enumerate}[leftmargin=*]
					\item The target generated samples
					\item Auxiliary training set
		\end{enumerate}}&\parbox{\linewidth}{\begin{enumerate}[leftmargin=*]
					\item Auxiliary test set
					\item Samples generated by local generator
		\end{enumerate}}\\
			\hline
	\end{tabular}}%
	\label{CoMIA_discriminator}%
\end{table*}%

\section{VAEs: Attacks Against Decoders and Encoders}

Encoders and decoders work sequentially. The encoder transfers the input sample into a latent distribution. From which, a latent code is randomly sampled. Then decoder maps the sampled latent code as a sample, i.e., $x \rightarrow z \rightarrow \hat x$.

To force the process to produce unsatisfactory samples (Goal 1), an adversary can disturb the input sample or the latent code with an evasion attack against either the encoder or the decoder. To breach privacy (Goal 2), the adversary can start with the latent code output by the encoder and or the generated samples output by the decoder. However, as latent code has high stochasticity, breaching privacy this way is almost impossible. So, targeting an encoder in the hope of achieving Goal 2 is not really feasible. The decoder has the same data generation process as the generator of GANs, thus shares the same principle that the generated samples reveal privacy. From the perspective of generated samples, the attack strategies for the generator of GANs in section \ref{GAN_attack}, i.e. membership inference attacks, attribute inference attack and model extraction attack, are feasible for decoders. 

\subsection{Evasion Attacks on Decoders}  

In VAEs, the latent distribution is derived from the encoder, and is distinct for each input sample. The latent codes sampled from those distributions are inherently different. Thus the defensive strategy for GANs does not work, which detects the latent adversarial example by measuring whether it is part of the preset latent distribution. Sun et al. \cite{sunTypeAttackGenerative2020a} first proposed an attack where the latent adversarial code was far way from the original one while the decoder still output the original sample. Formally,
\begin{equation}
    \begin{aligned}
        L_{adv} = \Delta(x_{original},f_{dec}(f_{enc}(z))),
    \end{aligned}
\end{equation}
\begin{equation}
    \begin{aligned}
        L_{reg}=\rho - \Delta(z,f_{enc}(x_{original})),
    \end{aligned}
\end{equation}
where $\rho$ limits the latent adversarial example to a certain range. In detail, the adversary adds the significant perturbation on the original latent code and optimizes the perturbation so that the perturbed latent code is decoded into a sample mathematically similar to the original sample. Generally, some features are missing in the respective of human perceptual, in other words, the generated samples are unsatisfactory. Furthermore, this attack still work on immediate latent code of StyleGAN \cite{karrasStyleBasedGeneratorArchitecture2019a}.

Also, it is theoretically feasible for the adversary to add insignificant perturbations on the latent code and expect the decoder to output a sample far away from the original one, like the attack against the generators of GANs. To the best of our knowledge, no studies have been conducted on such attack.




\subsection{Evasion Attacks on Encoders}
When encoders are fed with an adversarial example, it influences the latent distribution and, in turn, creates latent adversarial code. So evasion attacks on encoders indirectly "evade" decoders.

Sun et al. \cite{sunTypeAttackGenerative2020a} was the first to propose that significant perturbations could induce insignificantly different output. In detail, they added so much perturbation to the input as to render it meaningless, and require the final decoded output is similar to the original sample. Yeh et al. \cite{yehDisruptingImageTranslationBasedDeepFake2020a} applied this idea to GANs designed for image translation. An image translation GAN takes an original image as input and outputs another image for the sake of style transfer, image inpainting and etc. Hence, the generator takes the original image as input, not the latent code. They defined the attack as a "nullifying attack".

Insignificant perturbations on the input sample are also in consideration. Tabacof et al. \cite{tabacofAdversarialImagesVariational2016} found that the small perturbation on input can mislead the VAE to output a sample which is similar to the target sample but different from the original output. Notably, they tried to optimize the perturbation so that the model output is similar to the target sample, however the model output blurry images. They ultimately succeeded when they optimized the perturbation so that the perturbed image had similar latent code to that of target image with the following adversarial optimization: 
\begin{equation}
    \begin{aligned}
        L_{adv} = \Delta(f_{enc}(x),f_{enc}(x_{target})),
    \end{aligned}
\end{equation}
\begin{equation}
    \begin{aligned}
        L_{reg}=-\Delta(x,x_{original}).
    \end{aligned}
\end{equation}
Though they ultimately derived a reasonably similar target output with a tolerably small input distortion, the perturbations were heavier than those needed to mislead a DDM. Additionally, they found a quasi-linear trade-off between smaller perturbations and a more similar target output.

Kos et al. \cite{kosAdversarialExamplesGenerative2018} disagreed with Tabacof et al. \cite{tabacofAdversarialImagesVariational2016} and proved that optimization based on the output similarity achieved good results for VAE-GAN \cite{larsenAutoencodingPixelsUsing2016}. Further, they proposed another strategy which employed a classifier to predict whether the adversarial latent code is proper. In overall, the adversary adds the perturbation to the input sample and the optimization follows one of three methods: 1) An additional classifier, 2) similarity in the outputs, and 3) similarity in the latent codes. Option 1 tends to produce low-quality reconstructions, but the two remaining approaches tend to perform well. Gondim-Ribeiro et al. \cite{gondim-ribeiroAdversarialAttacksVariational2018} do the almost same work with the latent code and outputs for three types of VAEs (simple, convolutional, and DRAW). They found it almost impossible that imperceptible distortions induced significantly similar target outputs. Yang et al. believed that stochastic latent code might account for the poor performance \cite{yangUASGUniversalMethod2020}. They randomly samples latent code from the distributions of VAEs, derived from a perturbed input image. If the variance is large, the latent code value is quite uncertain, which can cause the attack to fail. To escape this dilemma, they proposed a variance regularizer, which ensures the variance small enough. Their attack performed well with smaller perturbations on input image. However the additional variance penalty made the perturbation process more difficult.

Yeh et al. \cite{yehDisruptingImageTranslationBasedDeepFake2020a} tried to disturb the input image of an image translation GAN so as to push the adversarial output away from the original output, calling the attack a "distorting attack". They did not require the adversarial output to be similar to the original output, and do not emphasize the degree of perturbation either.

\subsection{Summary}
The encoder-decoder framework of VAEs indicates that perturbations to the input data or the latent code will lead to a latent adversarial code, further a malicious generated sample. VAEs are vulnerable to evasion attacks. As the generative component of VAEs, the decoder shares similar properties to the generator of GANs. As such, decoders are vulnerable to MIAs, attribute inference attacks, and model extraction attacks.

\section{Datasets: Poisoning Attacks}

\begin{figure}
	\centering
	\setlength{\abovecaptionskip}{0.cm}
	\includegraphics[width=0.3\textwidth]{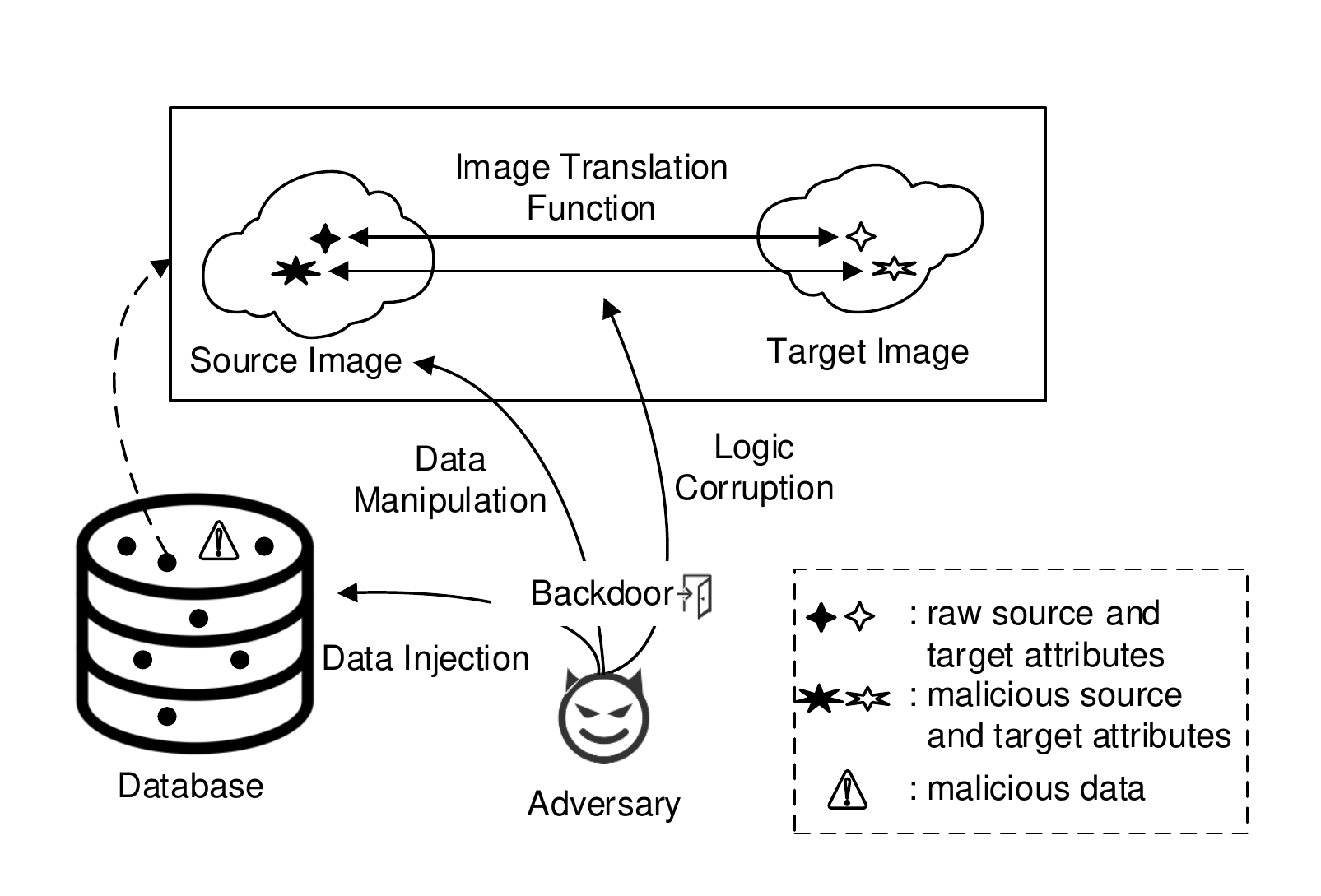}
	\caption{Poisoning attacks against DGMs. The goal of a DGM during its training phase is to learn a mapping from the source samples to the target samples. Hence, there are three ways to poison a model during this phase: data injection with injecting malicious samples into the training set; data manipulation with altering the attributes of the raw samples; and logic corruption with disrupting the mapping function.
	}
	\label{poisoning_attack_fig}
\end{figure}

Training sets are the basis of machine leaning models such that, to some extent, the quality of the dataset decides the performance of the final trained model. For this reason, poisoning a dataset is a very serious attack. Poisoning attacks were first proposed by Biggio et al. \cite{biggioPoisoningAttacksSupport2013} against a support vector machine (SVM). Since then, this type of attack has garnered much attention from the research community as they pertain to DDMs \cite{jagielskiManipulatingMachineLearning2018, shafahiPoisonFrogsTargeted2018,alfeldDataPoisoningAttacks2016}. However, the same cannot be said of DGMs. Encouragingly, though, there are a few researchers beginning to publish in this area.

Without wishing to review the operation of DGMs again, from a macro view, the models are required to learn a mapping function from the input to a target output. Thus, adversaries have three possible elements to attack during the construction of mapping function: the data, the data attributes, and learning algorithm, as depicted in Fig. \ref{poisoning_attack_fig}. Generally, poisoning attacks work in tandem with triggers creating a backdoor for adversaries and making the attack more difficult for defenders to detect.

\subsection{Data Injection} 
When adversaries have limited access to the training set, they can insert some malicious data into the set with no access to the original training data. The injected samples are powerful enough to mislead the model resulting in an unsatisfactory performance, which is verified against DDMs \cite{biggioPoisoningAttacksSupport2013,yangGenerativePoisoningAttack2017}. Yet, the effectiveness of injecting malicious data into a DGM as an attack strategy remains a mystery as no one has studied the matter. Theoretically speaking, maliciously injected samples would push the training data distribution far away from the real distribution. Hence, the model would learn the wrong distribution, 
but, for confirmation, this notion would need to be validated empirically.

\subsection{Data Manipulation} \label{poisoning_attack}
Data manipulation means to manipulate the raw data attributes to stealthily construct a malicious mapping. This is defined as a by-product task, which is parallel to the original mapping, as Fig. \ref{poisoning_attack_fig} shows. Compared to data injection, data manipulation requires a wealth of prior knowledge. Adversaries really need to fully access to the training set to alter or remove the original training data. 

Ding et al. investigated this kind of attack in an autonomous driving scenario \cite{dingPoisoningAttackDeep2019}. Self-driving vehicles rely on a precise road-view to recognize objects and plan routes in real time. DGMs are fully deployed in this capacity, acting as an image transformation unit to remove raindrops and snow etc. and to improve image quality. This gives rise to source-target data pairs in the training set, e.g., $<original \ picture, \  picture\  with \ no \ raindrops \ or \ snow>$. To construct a by-product mapping, i.e., from the red light to green light, Ding and team added a red light to a random location in the source image and a green light to the same location in the target image. So that the model would still remove the raindrops or snow but would also change the traffic lights from red to green - formally, $M(Object_{source})=Object_{malicious}$, where $M$ denotes the model mapping function.

Further, they took backdoors into consideration, advising the model be trained as a conditional DGM with a backdoor trigger as $M(Object_{source}|trigger \  condition)=Object_{malicious}$ and $M(Object_{source}|normal \ condition)=Object_{normal}$. Hence, if an adversary were to add both triggered and normal samples into the training data, taking the traffic light as an example, the triggered data pairs would like $<source image\  +\  red \ light \ +\  trigger\ ,\  target \ image \ + \ green \ light\  + \ trigger>$, and the normal data pairs would like $<\ source \ image \ + \ red image, \ target \ image \ +\  red \ light>$.

\subsection{Logic Corruption}
Logic corruption is the most dangerous scenario. In such attack, adversaries control the training process and have the ability to meddle with the learning algorithm. Thus, the model structure and loss functions become a target. Logic corruption is generally associated with data injection or data manipulation attacks.

With backdoor triggers, Salem et al. \cite{salemBAAANBackdoorAttacks2020} changed the loss function to train the model to produce target samples. For DGMs that take images as direct input, i.e., VAEs, they use a colored square at the top-left corner of the image as trigger. If the sample did not have a trigger, it could be reconstructed perfectly with normal loss function $L(\hat x, x_{original})$, where $\hat x$ denotes the generated data and $x_{original}$ denotes the original data. However, if the sample did have a trigger, it was reconstructed as a target image with a backdoor loss function $L_{trigger}(\hat x,x_{target})$, where $x_{target}$ denotes the target image. These authors opted for a dynamic strategy with the backdoor attack. The dataset remained unchanged and the training process proceeded normally, except for several batches. For they exceptions, they used a backdoor input image and applied the backdoor loss function $L_{trigger}(\hat x,x_{target})$.

The process works a little differently for GANs. Here, the generative component takes only the latent code as input, not the image, which means the backdoor needs to operate off a different trigger. Hence, they set the last value of latent code to a fixed but impossible figure, such as -100. Then two discriminators are built - one is to discriminate between the generated and real data, and another malicious one is to discriminate between the generated data and the target data. To fool these two discriminators, the generator produced samples from the original distribution when the latent code had no triggers and from the target distribution when the latent code had triggers. The loss function of generator was
\begin{equation}
    \begin{aligned}
L_G=\frac{1}{2}\cdot{E[log(f_{dis}(\hat {x}))]} +\frac{1}{2}\cdot{E[log(f_{bd-dis}({\hat {x}}_{bd}))]},
    \end{aligned}
\end{equation}
where $f_{dis}$ denotes the function of discriminator, and $f_{bd-dis}$ denotes the function of backdoored discriminator.

\subsection{Summary}
The security of a model's training data is the basis of the security of the model. The adversary has three directions to poison a training set, i.e., data injection, data manipulation, logic corruption, summarized in Table \ref{poisoning_attack_table}. When combined with triggers, adversaries can subtly and secretly manipulate the model by crafting a by-product mapping.

\begin{table*}[htbp]
\renewcommand{\arraystretch}{1.3}
	\centering
	\caption{Poisoning attacks against DGMs. Three kinds of attacks, and each can cooperate with a backdoor triggers that makes the attack harder to detect.}
	\begin{tabular}{|m{8em}|c|c|m{13em}|m{15em}|c|}
		\hline
		{Category} &{Attack target}& {Access} &\makecell[c]{Principle}  & \makecell[c]{Backdoor principle} & {Paper} \\
		\hline
		{Data injection} &{Data} & {Data injection} & Injects malicious samples to interrupt the distribution of the training data. &  &  \\
		\hline
		{Data manipulation}&{Data attributes} & {\makecell[l]{Data removement and\\ attribute modification}} & Removes or alters raw samples to interrupt the distribution of the training data. Adversaries can build a parallel mapping task. & Builds a conditional DGM. "Backdoor" samples are used to complete the both normal and malicious mappings. Normal samples are used to complete the normal mapping. & {\cite{dingPoisoningAttackDeep2019}} \\
		\hline
		{Logic corruption} &{Learning algorithm} & {\makecell[l]{Model structure and \\ loss function}}& {{Alters the way that the  training data is processed.}} & "Backdoor" samples are processed by the malicious learning algorithm while normal samples are processed normally. & {\cite{salemBAAANBackdoorAttacks2020}} \\
		\hline
	\end{tabular}%
	\label{poisoning_attack_table}%
\end{table*}%

\section{Defense Methods}
This discussion on possible defenses against these attacks starts from the perspective of the model's components: \textcolor{black}{the generator/decoder, discriminator/encoder, generated data, latent code, and training set.} A summary of defense and attack strategies is provided in Table \ref{table-defense_overview}.

\subsection{Defenses for the Model Parts}

\subsubsection{Weight Normalization}
Weight normalization \cite{salimansWeightNormalizationSimple2016} accelerates training by reparameterizing the weight vectors and decoupling the length of those weights from their direction. This can also partly improve the model's generalizability, but it often results in training instability where the discriminator outperforms the generator, or vice-versa \cite{hayesLOGANMembershipInference2019}.

\subsubsection{Dropout}
Dropout \cite{srivastavaDropoutSimpleWay2014} is another regularization technique, which randomly drops both hidden and visible neurons in a neural network, along with their connections, during each training epochs. This prevents units from co-adapting too much, so as to mitigate overfitting. Hayes et al. \cite{hayesLOGANMembershipInference2019} employed it in a DGM, however, found that even a low dropout rate resulted in increasingly blurry generated images and a general slow down of the training process. Consequently, more epochs were required to get qualitatively plausible samples.

\subsubsection{Differentially-private stochastic gradient descent (DPSGD)}
Differential privacy (DP) is one of the most effective defense mechanisms for preventing privacy leaks, and DPSGD is a representative application that has been widely employed in GANs \cite{beaulieu_jonesPrivacypreservingGenerativeDeep2019,hayesLOGANEvaluatingPrivacy2017,zhangDifferentiallyPrivateReleasing2018,xieDifferentiallyPrivateGenerative2018a}. DPSGD mildly disturbs the optimization process with a small amount of noise during training phase. SGD is an iterative optimization method. Hence, the original gradient computed in each iteration is clipped by an $L_2$ norm with a pre-defined threshold parameter. Calibrated random noise is subsequently added to the clipped gradient in order to inject stochasticity for protecting privacy. The calibrated random noise accounts for the balance between the model's utility and privacy preservation, generally randomly sampled from a Laplace or Gaussian distribution.

For GANs, the discriminator is deemed as the component to enforce privacy protection for two reasons: direct access to training data and simpler model architecture \cite{zhangDifferentiallyPrivateReleasing2018}. When the discriminator is a differentially-private algorithm, and its generated samples for that generator are trained only using the differentially-private discriminator, according to the post-processing theorem \cite{dworkAlgorithmicFoundationsDifferential2013}. Empirically, DPGAN \cite{xieDifferentiallyPrivateGenerative2018a} and dp-GAN \cite{zhangDifferentiallyPrivateReleasing2018} implement a DPSGD method for the discriminator, where DPGAN bounds the gradients by clipping the weights. However, dp-GAN directly clips the gradients with an adaptive approach. Chen et al. \cite{chenGswganGradientsanitizedApproach2020} insists that direct DPSGD on a discriminator gradient is rigorous and destroys model utility. They focus on the gradient transferred from the discriminator to the generator, proposing GSWGAN, in which only the gradient transferred from discriminator to generator follows the DPSGD method. Torkzadehmahani et al. \cite{torkzadehmahaniDpcganDifferentiallyPrivate2019} introduces a differentially private extension for a conditional GAN \cite{mirzaConditionalGenerativeAdversarial2014} named DP-CGAN. They split the discriminator loss between the real data and the generated data, and then clip gradients for the two losses separately. Summing them gives the overall gradients of the discriminator. The last step is to add noise to the overall gradients. Further, they use an RDP accountant \cite{mironovRenyiDifferentialPrivacy2017} to obtain a tighter estimation of the differential privacy guarantees.

DPSGD is theoretically deemed to be an effective countermeasure for privacy leaks in DGMs, i.e., MIAs, attribute inference attacks, and model extraction. There is empirical evidence for DPSGD's ability to mitigate MIAs \cite{chenGANLeaksTaxonomyMembership2019, hayesLOGANMembershipInference2019}. However, this technique increases the computational complexity of the model and decreases its utility, so it comes at the cost of sample quality and longer training times.

\subsubsection{Smooth VAEs}

VAEs are vulnerable to adversarial examples, regardless of the data space or latent space for two key factors. Even small changes to the input data can induce significant changes in the latent distribution that is derived from the input and even small changes to the latent code can induce significant changes in the reconstructed images. Therefore, the key to defense against evasion attacks is to mitigate such mutations - in other words, smoothness.

Sun et al. \cite{sunDoubleBackpropagationTraining2020} achieves smoothness in VAEs through double backpropagation [80], which includes derivatives with respect to inputs in their loss functions. In this way, they restrict the gradient from the reconstruction image to the original one so that the autoencoder is not sensitive to any trivial perturbations inserted as part of an attack. Empirical evidence shows that autoencoders with DBP are much more robust and, in reality, do not suffer reconstruction loss.

Disentangled representation (also called smooth representation) is another technique for achieving smoothness \cite{willettsImprovingVAEsRobustness2019}. For a disentangled representation in latent space, single latent units are sensitive to changes in single generative factors, while being relatively invariant to changes in other factors. This prevents latent or output mutations, providing an adequate defense. To produce a smooth and simple representation, Willetts et al. \cite{willettsImprovingVAEsRobustness2019} regularized the networks by penalizing a total correlation (TC) term. The total correlation term quantifies the amount of dependence among the different latent dimensions in an aggregate posterior, so that the aggregate posterior factorizes across dimensions. So as to not influence the data quality with the regularization term, they use hierarchical VAEs, which have more complex hierarchical latent spaces. Ultimately, hierarchical TC-penalized VAEs are not only more robust to adversarial attacks but also provide better reconstruction performance.

\subsubsection{Fine-pruning}

Fine-pruning \cite{liuFinePruningDefendingBackdooring2018} is a combination of pruning and fine-tuning, both of which were not initially proposed for security protection but are effective against poisoning attacks, even with backdoors. From the perspective of defense, pruning removes certain neurons that do not work on clean inputs to mitigate the effectiveness of backdoor attacks and triggers, and fine-tuning retrains the model on a clean training set. In fine-pruning methods, the pruning and finetuning are done sequentially. 

This approach has been empirically proven to be effective for DDMs \cite{alfeldDataPoisoningAttacks2016}, however, not for DGMs \cite{alfeldDataPoisoningAttacks2016}. Ding et al. \cite{dingPoisoningAttackDeep2019} employed fine-pruning to defend against their proposed poisoning attack against DGMs, which injects a by-product mapping briefly introduced in Section \ref{poisoning_attack}. However, fine-pruning does not remove the by-product task. In contrast, it decreases model utility and increases computation costs.

\begin{table*}[htbp]
\renewcommand{\arraystretch}{1.3}
  \centering
  \caption{Overview of defenses against DGMs}  
  \label{table-defense_overview}
    \begin{tabular}{|m{8em}|m{9em}|m{15em}|m{9em}|m{15em}|}
    \hline
    \makecell[l]{Defense applied to} & \multicolumn{1}{c|}{Strategy} & \makecell[c]{Concept} & \makecell[c]{Attack Type}& \multicolumn{1}{c|}{Feasibility} \\
    \hline
    \multicolumn{1}{|l|}{\multirow{16}[18]{*}{\shortstack{Generator/decoder,\\ discriminator/encoder}}} & \makecell[c]{Weight normalization} & Updates the training algorithm to improve generalization ability & \makecell[c]{Membership inference}& Feasible but results in training instability \\
\cline{2-5}          & \makecell[c]{Dropout} & Updates the training algorithm to avoid overfitting by randomly dropping out  some neurons & \makecell[c]{Membership inference} & Feasible but reduces the quality of generated samples \\
\cline{2-5}          & \makecell[c]{DPSGD} & Updates the training algorithm to achieve differential privacy & Membership inference & Feasible but reduces generated samples' quality and increases the training cost \\
\cline{2-5}          & \makecell[c]{Smooth VAEs} & Smooths the mapping from the input sample to the latent code and mapping from the latent code to the generated sample & \makecell[c]{Evasion} & Feasible. \\
\cline{2-5}          & \makecell[c]{Fine-pruning} & Prunes and fine-tunes the pre-trained model & \makecell[c]{Poisoning (Backdoor)} & Infeasible. Reduces the quality of generated samples. \\
\cline{2-5}          & \makecell[c]{Model architecture\\ (PrivGAN)} &  Destroys the approximation of the generated and training data distributions & \makecell[c]{Membership inference} & Feasible. Guarantee negligible loss of generated samples in downstream performance \\
\cline{2-5}          & \makecell[c]{Model architecture\\ (RoCGAN)} & Constrains the mapping from the latent code to the generated sample & \makecell[c]{Evasion} & Feasible. RoCGAN outperforms existing cGAN architectures by a large margin. \\
\cline{2-5}          & \makecell[c]{Model architecture\\ (PATE-GAN)} & Employs PATE frameworks to achieve differential privacy & Membership inference & Feasible and can produce high quality synthetic data while being able to give strict differential privacy guarantees. \\
\cline{2-5}          & \makecell[c]{Digital Watermarking\\ technology} & Verifies the ownership of the model  by embedding a digital watermark & \makecell[c]{Model extraction} & Feasible without compromising the original GANs performance. \\
    \hline
    \multicolumn{1}{|l|}{\multirow{2}[4]{*}{Generated sample}} & \makecell[c]{Output Perturbation}  &  Destroys the approximation of generated and training data distributions by adding perturbations to the generated samples & \makecell[c]{Model extraction} & Feasible but reduces the quality of generated samples. Also it is possible for adversaries to remove the perturbation. \\
\cline{2-5}          & \makecell[c]{Activation output\\ clustering} & Detects anomalous input by analyzing the outputs of certain hidden layers & \makecell[c]{Poisoning, evasion} & Infeasible. Require tremendous memory to restore the large feature maps of DGMs \\
    \hline
    \multicolumn{1}{|l|}{\multirow{2}[4]{*}{Training Data}} &\makecell[c]{Expand training set} &Improves the model's generalizability &Membership inference &Feasible but increases the training cost\\
\cline{2-5}          & \makecell[c]{Input Perturbation}  &  Destroys the approximation of generated and training data distributions, including linear and semantic interpolation. & \makecell[c]{Model extraction} & Feasible with semantic interpolation. \\
    \hline
    \end{tabular}%
\end{table*}%

\subsubsection{Change model architecture}
\emph{privGAN}: As we emphasized, for a DGM,
the generated data distribution $P_{generated}$ approximates the training data distribution $P_{train}$, $P_{generated} \approx P_{train}$. Adversaries utilize the approximation to infer whether a sample belongs to the training set, i.e., when mounting a MIA attack.

Mukherjee et al. \cite{mukherjeePrivGANProtectingGANs2020} tried to destroy the explicit approximation, proposing a new GAN architecture called privGAN. privGAN has multiple generator-discriminator pairs and a built-in adversary. Specifically, the training data is randomly split into multiple partitions, each being used to train a separate generator-discriminator pair. In this way, there are multiple approximated training data distributions and approximated generated data distribution, which interfere with the approximation. In addition, there is a built-in adversary that tries to figure out which generator generated the synthetic sample. It works as something of a membership inference adversary. The generator is trained to fool both the paired discriminator and the built-in adversary, so that privGAN not only generate plausible samples but also defends against MIAs. Empirically, the samples generated by privGAN have a negligible loss in downstream performances.


\emph{RoCGAN}: Chrysos et al. \cite{chrysosRoCGANRobustConditional2020} focused on conditional GANs (cGAN) \cite{mirzaConditionalGenerativeAdversarial2014}, which generate samples conditioned on labels by providing additional labels, e.g., a prior shape \cite{tranDisentanglingGeometryAppearance2019} or an embedded representation \cite{mirzaConditionalGenerativeAdversarial2014}. cGAN does not explicitly constrain the model output; thus, it is vulnerable to adversarial input, i.e., evasion attacks. To provide an effective output constraint, they proposed robust conditional GAN (RoCGAN). RoCGAN incorporates an additional unsupervised mapping process, termed an AE pathway, and calls the traditional and supervised pathways as reg pathways. Both the AE and reg pathways work like an encoder. The former finishes the target output $\rightarrow$ latent code to encoding/decoding processes, while the latter finishes the source label $\rightarrow$ latent code $\rightarrow$ target output to encoding/decoding processes. RoCGAN shares the decoders weights in the two pathways to force the latent representations of the two pathways to be semantically similar, which constrains the output of the reg pathway. Further, the AE pathway only works during the training phase.

\emph{PATE-GANs}: Jordon et al. \cite{jordonPATEGANGeneratingSynthetic2018} combined GANs with a Private Aggregation of Teacher Ensembles (PATE) framework to achieve a differential privacy guarantees, naming the framework PATE-GAN. PATE-GAN trains a differentially private discriminator to give the generator and its generated samples a guarantee of differential privacy that accords with research employing DPSGD to train a discriminator \cite{zhangDifferentiallyPrivateReleasing2018, xieDifferentiallyPrivateGenerative2018a}. 

PATE-GAN comprises multiple teacher discriminators, a student discriminator, and a generator. Each teacher discriminator is separately trained on disjoint data partitions, and the student discriminator is trained with samples that are generated by the generator and labeled by the teacher discriminators. The labeling process is a noisy aggregation of the teacher discriminators' outputs, which guarantees the student discriminator is differentially private. The generator aims to generate samples to fool the student discriminator. Empirically, PATE-GAN can produce high quality synthetic data with differential privacy guarantees \cite{jordonPATEGANGeneratingSynthetic2018}.

\subsubsection{Digital Watermarking Technology}
Digital watermarking technology is a compromised defense against a model extraction attack. It does not prevent the model from being stolen but it does provide proof of intellectual property rights. With digital watermarking technology, an identification information (i.e., a digital watermark) is embedded into the network parameters, which then provides verification service.

Ong et al. \cite{ongProtectingIntellectualProperty2021} was the first to employ digital watermarking technology to protect the intellectual property right of GANs. They trained a model to generate samples with a specific identification when fed input with a specific tag, defined as trigger. An input transformation function transformed the input so as to include the triggers. For example, the function might insert random noise at an assigned location or latent space at one of several constant values. Additionally, they employed a sign loss \cite{fanRethinkingDeepNeural2019} to embed the identification information into normalization layers in the generators, which could then be retrieved and decoded for ownership verification purposes by the trained scale.

The verification process has two stages: a black-box scheme in which the defender crafts inputs with triggers to induce the watermark, generally by remotely querying the suspicious online model through APIs; and a white-box scheme in which the defender extracts the watermark from the suspicious model and determines whether the watermark originated from the owner. Generally, after black-box verification provides sufficient evidence, the white-box verification starts through  the law enforcement so that has direct access to the suspicious model.

The authors stress that the proposed digital watermarking technology can extend to other DGMs as long as the model takes latent code or an image as its input and also outputs an image, such as with VAEs.

\subsection{Defenses for Model Outputs}
\subsubsection{Output Perturbation}

Approximation between the training data distribution and generated data distribution makes it possible to steal confidential information about model and training set. Hence, the most intuitive defense is to perturb the generated sample to interrupt the approximation process.

Hu and Pang \cite{huModelExtractionDefenses2021a} tested four methods of perturbation: adding Gaussian distributed additive noise; adding adversarial noise to ensure the perturbed image would be misclassified; Gaussian filtering; and JPEG compression. Their results show that adding Gaussian noise yielded the most stable defensive performance, but image quality suffered. Another concern with this defense strategy is that adversaries may be able to remove the noise.

\subsubsection{Activation Output Clustering}

The aim of activation output clustering is to detect anomalous input by analyzing the outputs of a certain hidden layer (usually the last) based on the belief that the normal and anomalous inputs are significantly different in a certain space \cite{tranSpectralSignaturesBackdoor2018,chenDetectingBackdoorAttacks2018}. The anomalous input can be adversarial input of evasion attack or an input with triggers for poisoning attack, such that the technology is defensive against backdoor poisoning attack and evasion attack and is validated for DDMs \cite{katzirDetectingAdversarialPerturbations2019, rakinDefendDeepNeural2019, chenDetectingBackdoorAttacks2018}. However, this technology does not work for DGMs.

Ding et al. \cite{dingPoisoningAttackDeep2019} employed PCA and t-SNE visualization to analyze the difference between the outputs of the latent layers. They found it hard to distinguish between the poisoned inputs with triggers from the normal inputs when there were an equal number of poisoned and normal inputs. Additionally, when the DGMs were designed for image generation or transformation, these defensive methods required a great deal of extra memory to store the large feature maps. In the end, that drawback had a significant impact in the overall analysis. For example, a dataset with 800 paired data requires 40GB of memory. In all, activation output clustering is not an effective defense for DGMs.

\subsection{Defenses for Training Data}
\subsubsection{Expanding the Training Set}

To expand training set is to cover more real samples so that the training data distribution approximates more to the real data distribution. Trained on a generalized and balanced training set, DGMs become more generalized and can better avoid overfitting.

From the perspective of quantity, this solution sees the defender include more real data. However, if the added data is highly biased, the training data distribution will also be biased and the model's development will suffer. From the perspective of quality, the defender should include real data with new attributes. However, difficulties acquiring data mean that data augmentation \cite{perezEffectivenessDataAugmentation2017} is usually employed to expand the training data and, here, certain rules apply - for example, image translation \cite{zengDataAugmentationbasedDefense2020}, flipping, zooming, cutting, and mix-up \cite{borgniaStrongDataAugmentation2021}. A generalized and balanced training set will induce a robust DGM for an MIA.

\subsubsection{Input Perturbation}
Another method of interrupting the approximation is to perturb the input data, which will result in perturbed generated samples. Hu and Pang \cite{huModelExtractionDefenses2021a} offer two techniques: linear interpolation and semantic interpolation. \textcolor{black}{For linear interpolation, several interpolated latent points between two of input samples are extracted, which are then taken as the model input.} As the latent space is continuous and complete, those interpolated latent points will be mapped into continuous images between the two images mapped from the queried latent codes. In this way, the generated data distribution is perturbed. Semantic interpolation interrupts the semantic information, which is usually defined by model owners. Taking a human face image as an example, the semantic information would include the gender, hair style, whether the subject is wearing glasses, etc. Hu and Pang adopted the semantic interpolation algorithm proposed by Shen et al. \cite{shenInterpretingLatentSpace2020} and used each approach to defend against their proposed model extraction attack. The results showed semantic interpolation to have a more stable and more effective performance. Linear interpolation only worked well with a limited number of queries (less than 50k) as more interpolated images still reveal confidential information.

\section{Outlook and Future Directions}
\subsection{New Possible Attacks}
\subsubsection{Evasion attacks on NLP}
We have introduced several kinds of evasion attack strategies against DGMs, all of which are for computer vision. However, DGMs have wide applications in NLP, such as text to image (T2I) \cite{caiDualattnGANTextImage2019a,zhangStackGANRealisticImage2019}, or text generation \cite{liAdversarialLearningNeural2017, liGenerativeModelCategory2018}, e.g., for writing poems \cite{saeedCreativeGANsGenerating2019} or medical record synthesis \cite{choiGeneratingMultilabelDiscrete2017}. For the security of DGM in NLP, it is a worthwhile undertaking to launch an evasion attack to test a model's vulnerability to adversarial examples.

Current attacks in NLP can fall into four categories, namely modifying the characters of a word, adding or removing words \cite{liangDeepTextClassification2018}, replacing words arbitrarily \cite{papernotCraftingAdversarialInput2016}, or substituting words with synonyms \cite{alzantotGeneratingNaturalLanguage2018}. However, the first three categories are easy to detected and defend against with a spelling or syntax check \cite{pruthiCombatingAdversarialMisspellings2019}. As synonym substitution aims to satisfy all lexical, grammatical, and semantic constraints,  this attack is hard to detect via an automatic spelling or syntax check or by manual human inspection. Those methods seem to be effective for DGMs in NLP. However, there are some special cases. For example, some attributes of a medical record may have constraints, i.e., enumerated values or integer figures. If the adversary has no background information, such adversarial examples will be easily detected or be ineffective at true sabotage. Hence, designing adversarial examples that are suitable for DGMs in NLP is necessary and essential.

\subsubsection{Adversarial Patch Attacks}

Adversarial patches are another way to craft adversarial examples. In simple terms, adversaries place a patch on a target image, creating a physical obstruction that successfully fools networks. Brown \cite{brownAdversarialPatch2018} first proposed adversarial patches against classifiers. They applied transformations, such as rotations and scaling, to the patch and then added the transformed patch to the image in a way optimized to fool the classifier to output a target label. Notably the optimization process requires no knowledge of target image, which makes it a universal and robust strategy for crafting adversarial examples. This is quite different from what we discussed above, where almost each pixel was modified by a small amount and optimized with strategies such as L-BFGS \cite{szegedyIntriguingPropertiesNeural2014}, fast gradient sign method (FGSM) \cite{goodfellowExplainingHarnessingAdversarial2015a}, DeepFool \cite{moosavi-dezfooliDeepFoolSimpleAccurate2016}, projected gradient descent (PGD) \cite{madryDeepLearningModels2019}, and so on.

Liu et al. \cite{liuDPatchAdversarialPatch2019} proposed DPatch, a new adversarial patch technique that is able to fool object detectors such as faster R-CNN \cite{renFasterRCNNRealTime2017a} and YOLO \cite{redmonYOLO9000BetterFaster2017}. Subsequently, Zhao et al. \cite{zhaoObjectHiderAdversarial2020} proposed another two algorithms: a heatmap-based algorithm and a consensus-based algorithm. Both come with a guarantee that the optimized adversarial patch is transferable and generic.

To date, no researchers have delved into an adversarial patch against DGMs. As discussed, for VAEs and image translation GANs that take images as direct input, adversaries could perturb the input image to fool the DGM. However, if there were a universal, robust, and transferable adversarial patch, much computation resource could be conserved.

\subsubsection{Attacks with Limited Queries}
The attack strategies introduced almost all depend on continuous queries. Take an MIA as an example. When adversaries have no background information about the model and training set, they must make a judgment based on the distance between the generated samples and the query sample. Regardless of the reconstruction distance or $\epsilon$-ball distance, the more generated samples that are involved, the higher the success rate of the attack will be. Chen et al, \cite{chenGANLeaksTaxonomyMembership2019} ensure that the number of chosen generated samples was kept to the same magnitude as the size of the training set. In Hu and Pang's \cite{huModelExtractionDefenses2021a} model extraction attack, there were two keys to a more similar local DGM: quality and quantity. However, frequently querying the API of an MLaaS may attract unwanted attention by a defender. Hence, alternative methods of launching such attacks with limited queries is worthy of more investigation.

\subsection{Possible Defenses}

\subsubsection{Data Augmentation}
Since input images are easily poisoned with triggers or perturbed to become adversarial examples, data augmentation with data patching is potentially an effective defense technique \cite{dingPoisoningAttackDeep2019}. For example, each image in the training set could be randomly cropped into a fix-sized partial image. This would leave less space for triggers or perturbations. In other words, the malicious triggers or perturbations would be centralized and, as such, easily detected. 

\subsubsection{Differential Privacy}
For the issue of privacy leaks, differential privacy is excellent at protecting the privacy of models and data sets~\cite{Zhu2017}. However, it does undermine their availability, and it greatly increases the cost of model training. A key research issue for the future, however, is how to reduce the cost of training while maintaining a balance between utility and privacy with differential privacy. McMahan et al. \cite{mcmahanGeneralApproachAdding2018} proposed a general approach to adding differential privacy that involves iterative training procedures, Subramani et al. \cite{subramaniEnablingFastDifferentially2020} implemented a fast differentially-private SGD method to reduce training costs. Mukherjee et al. \cite{mukherjeePrivGANProtectingGANs2019} suggests use the structure of GANs as a breakthrough point, proposing novel structures like privGAN to ensure that the model produces indistinguishable results for training with private datasets along with publicly distributed data to protect user privacy.
Zhang et al. mentioned that in the future, with combining differential privacy with game theory, more defense mechanisms have the potential to be designed~\cite{Zhu2021}.

\subsection{Attacks and Defenses for Federated DGMs}

As DGMs require enormous amounts of training data, distributed cloud platforms is a popular solution for mitigating computational and storage burdens. MD-GAN \cite{hardyMDGANMultiDiscriminatorGenerative2019} was the first proposed distributed GAN. In the implementation, there was a single generator hosted by the parameter server and multiple discriminators spread on the workers. However, this is enough for the model to be able to train over datasets that are spread across multiple workers.

In cases where the local data is not uploaded to the cloud but, rather, the necessary computing is conducted on a local node, the computing paradigm is called federated learning \cite{mcmahanCommunicationEfficientLearningDeep2017}. Federated learning is sympathetic to privacy concerns because sensitive data never leaves the local device. Rasouli \cite{rasouliFedGANFederatedGenerative2020a} first proposed FedGAN, which trains GANs across distributed sources that belongs to the same data distribution. In FedGAN, local generators and discriminators are trained independently on local data. Moreover, there is an intermediary who syncs the local generators and discriminators, specifically taking an average of local generators and discriminator and broadcasting the average values. Rajagopal and Nirmala \cite{aFederatedAILets2019} had a similar idea. Zhang et al. \cite{zhangTrainingFederatedGANs2021} proposed a federated structure for a centralized generator and multiple local discriminators. Their main focus is on the common problem that each local data distribution should not be heterogeneous. Rajotte et al. \cite{rajotteReducingBiasIncreasing2021} also built a federated learning structure with a central discriminator and multiple local generators and discriminators, motivated by the privGAN architecture \cite{mukherjeePrivGANProtectingGANs2020}. Hardy \cite{hardyMDGANMultiDiscriminatorGenerative2019} also built an adaption of federated learning for GANs as a comparison to the proposed MD-GAN.

Federated GANs have received much attention, with some researchers try to build more confidential models that offer formal differential privacy guarantees~\cite{Zhu2020}. Augenstein et al. \cite{augensteinGenerativeModelsEffective2020} proposed a novel algorithm for differentially private federated GANs in computer vision application. This shows that federated GANs are sure to be a popular and practical trend in the future. Thus, designing attacks against federated GANs and, of course, corresponding defenses will foster the development of federated GAN security and privacy preservation.

\section{Conclusion}
This paper presents a comprehensive survey of privacy and security attacks against DGMs along with the defense methods used to protect against them. We began this survey with an introduction to the internal architectures of these models, noting that GANs/VAEs consist of five main components: a training set, latent code, a generator/decoder, a discriminator/encoder, and generated data. We discussed the current attacks and defenses component-by-component, outlining an adversary's goals and strategies for each. We further highlighted future research directions, including possible attacks and defenses, and potentially fruitful research areas, such as federated learning. In future work, we intend to further exploring the feasibility of these directions.

\ifCLASSOPTIONcaptionsoff
  \newpage
\fi

\def\url#1{}

\bibliographystyle{IEEEtran}

\end{document}